\pdfoutput=1
\documentclass[12pt]{article}
\usepackage[utf8]{inputenc}      
\usepackage[colorlinks=true,citecolor=blue,urlcolor=blue,linktocpage=true,linkcolor=blue]{hyperref}
\usepackage{amsmath,amssymb,mathtools}
\usepackage[T1]{fontenc}          
\usepackage{booktabs,tabularx}
\usepackage{graphicx}
\usepackage{xspace}
\usepackage[usenames]{xcolor}\definecolor{fscolor}{RGB}{44,118,255}
\usepackage{geometry}
\usepackage{todonotes}
\usepackage{listings}
\usepackage[absolute]{textpos}
\usepackage[many]{tcolorbox}
\usepackage{xparse}
\usepackage[font=small,labelfont=bf,format=plain,margin=0.05\textwidth]{caption}
\usepackage{bbm}
\usepackage{tabularx}
\usepackage{soul}
\usepackage[capitalize]{cleveref}
\usepackage{slashed}
\usepackage{subcaption}
\usepackage{pifont}
\usepackage[shortlabels]{enumitem}
\usepackage{listings}
\usepackage[frozencache=true]{minted}
\definecolor{bg}{rgb}{0.95,0.95,0.95}
\usepackage[sorting=none,%
  citestyle=numeric-comp,%
  bibstyle=mynumeric,%
  giveninits=true]{biblatex}

\usepackage[super]{nth}

\allowdisplaybreaks

\evensidemargin \oddsidemargin
\newcommand{\HiTo}{\texttt{HiggsTools}\xspace}
\newcommand{\HiBo}{\texttt{HiggsBounds}\xspace}
\newcommand{\HiSi}{\texttt{HiggsSignals}\xspace}
\newcommand{\HiPr}{\texttt{HiggsPredictions}\xspace}

\newcommand{\cpp}{\texttt{C++}\xspace}
\newcommand{\py}{\texttt{Python}\xspace}
\newcommand{\mat}{\texttt{Mathematica}\xspace}

\newcommand{\cp}{\ensuremath{\mathcal{CP}}\xspace}
\newcommand{\tev}{\,\, \mathrm{TeV}}
\newcommand{\gev}{\,\, \mathrm{GeV}}
\newcommand{\mev}{\,\, \mathrm{MeV}}
\newcommand{\fb}{\,\, \mathrm{fb}}
\newcommand{\fbinv}{\ensuremath{\mathrm{fb}^{-1}}\xspace}

\newcommand{\address}{\sl\small}

\newcommand\refse[1]{Sect.~\ref{#1}}

\newcommand\citere[1]{Ref.~\cite{#1}}
\newcommand\citeres[1]{Refs.~\cite{#1}}

\def\reffi#1{\mbox{Fig.~\ref{#1}}}

\NewBibliographyString{refname}
\NewBibliographyString{refsname}
\DefineBibliographyStrings{english}{%
  refname = {Ref\adddot},
  refsname = {Refs\adddot}
}
\DeclareCiteCommand{\ccite}
  {%
  \ifnum\thecitetotal=1
    \bibstring{refname}%
  \else%
    \bibstring{refsname}%
  \fi%
  \addnbspace\bibopenbracket%
  \usebibmacro{cite:init}%
   \usebibmacro{prenote}}
  {\usebibmacro{citeindex}%
   \usebibmacro{cite:comp}}
  {}
  {\usebibmacro{cite:dump}%
   \usebibmacro{postnote}%
   \bibclosebracket}
\newrobustcmd*{\Ccite}{\bibsentence\ccite}

\begin{document}

\thispagestyle{empty}
\def\thefootnote{\fnsymbol{footnote}}

\begin{flushright}
DESY-22-130\\
IFT--UAM/CSIC--22-123\\
EFI-22-9\\
KA-TP-25-2022
\end{flushright}
\vspace{2em}
\begin{center}
{\large{\bf
\HiTo: BSM scalar phenomenology with new versions of \\[.4em]
\HiBo and \HiSi
}}

\vspace{1em}
 {
Henning Bahl$^{1}$\footnotetext[0]{emails: hbahl@uchicago.edu, thomas.biekoetter@desy.de, Sven.Heinemeyer@cern.ch, cheng.li@desy.de,\\ \mbox{}\hspace{18mm} steven.paasch@desy.de, georg.weiglein@desy.de, jonas.wittbrodt@desy.de},
Thomas Biekötter$^{2}$
, Sven Heinemeyer$^{3}$
, Cheng Li$^{4}$
, Steven Paasch$^{4}$
,\\[.3em] Georg Weiglein$^{4,5}$
, and Jonas Wittbrodt$^{6}$
 }\\[2em]
 {\address $^1$    University of Chicago, Department of Physics and Enrico Fermi Institute,\\[0.2em] 5720 South Ellis Avenue, Chicago, IL 60637 USA}\\[0.2em]
 {\address $^2$    Institute for Theoretical Physics, Karlsruhe Institute of Technology,
Wolfgang-Gaede-Str.~1, 76131 Karlsruhe, Germany}\\[0.2em]
 {\address $^3$    Instituto de F\'isica Te\'orica, (UAM/CSIC), Universidad Aut\'onoma de Madrid,\\[0.2em] Cantoblanco, E-28049 Madrid, Spain}\\[0.2em]
 {\address $^4$    Deutsches Elektronen-Synchrotron DESY, Notkestr.~85, 22607 Hamburg, Germany}\\[0.2em]
 {\address $^5$    II.\  Institut f\"ur  Theoretische  Physik, Universit\"at  Hamburg, Luruper Chaussee 149, 22761 Hamburg, Germany}\\[0.2em]
 {\address $^6$    Department of Astronomy and Theoretical Physics, Lund University, Sölvegatan 14A, 223 62 Lund, Sweden\footnote{Former address.}}
\def\thefootnote{\arabic{footnote}}
\setcounter{page}{0}
\setcounter{footnote}{0}
\end{center}
\vspace{1ex}

\begin{abstract}\noindent
    The codes \HiBo and \HiSi compare model predictions of BSM models with extended scalar sectors to searches for additional scalars and to measurements of the detected Higgs boson at $125\gev$. We present a unification and extension of the functionalities provided by both codes into the new \HiTo framework. The codes have been re-written in modern \cpp with native \py and \mat interfaces for easy interactive use. We discuss the user interface for providing model predictions, now part of the new sub-library \HiPr, which also provides access to many cross sections and branching ratios for reference models such as the SM. \HiBo now implements experimental limits purely through \texttt{json} data files, can better handle clusters of BSM particles of similar masses (even for complicated search topologies), and features an improved handling of mass uncertainties. Moreover, it now contains an extended list of Higgs-boson pair production searches and doubly-charged Higgs boson searches. In \HiSi, the treatment of different types of measurements has been unified, both in the $\chi^2$ computation and in the data file format used to implement experimental results.

\end{abstract}
\setcounter{footnote}{0}
\renewcommand{\thefootnote}{\arabic{footnote}}

\newpage

\tableofcontents

\newpage


\section{Introduction}

With the discovery of a Higgs boson  with a mass of $\sim 125\gev$ at the LHC~\cite{ATLAS:2012yve,CMS:2012qbp}, the first (potentially) elementary scalar particle was observed. This discovery marks an important milestone in the quest to unravel the nature of electroweak symmetry breaking (EWSB). The further investigation of EWSB --- i.e., the precise determination of the properties of the Higgs boson at $125\gev$ as well as the search for additional scalar bosons --- is one of the main tasks of the LHC physics program.

Many models beyond the SM (BSM) contain extensions of the Standard Model (SM) Higgs-boson sector, thus predicting additional scalar particles. Well-known examples include the extension of the SM Higgs sector by additional $SU(2)_L$ singlets, doublets and also higher representations. The LHC searches carried out so far have not led to the discovery of additional scalar bosons. Correspondingly, the searches have resulted in exclusion limits constraining the parameter space of BSM models with extended scalar sectors.

Similarly, measurements of the properties of the Higgs boson at $125 \gev$ so far have not found any conclusive deviation from the SM predictions. In turn, also these measurements constrain the parameter space of BSM models which naturally predict modifications of the couplings of the Higgs boson at $125 \gev$ (commonly called the ``SM-like Higgs boson'') w.r.t.\ the corresponding SM predictions.

Consequently, every BSM model modifying the scalar sector of the SM --- either by modifying the couplings of the SM-like Higgs boson or by adding new BSM scalars to the theory --- should be tested against all the available data collected at the LHC, at LEP and other colliders. Due to the large number of available searches and measurements, checking the consistency of a BSM parameter point with these experimental results is not feasible without the development of dedicated computer tools to facilitate this task.

The codes \HiBo~\cite{Bechtle:2008jh,Bechtle:2011sb,Bechtle:2013wla,Bechtle:2020pkv} (see also \ccite{Bahl:2021yhk}) and \HiSi~\cite{Bechtle:2013xfa,Bechtle:2020uwn} have been developed in this spirit. While \HiBo allows checking BSM models against exclusion limits from searches for new scalar bosons, \HiSi allows one to check the compatibility of the model with the LHC rate measurements of the Higgs boson at $125\gev$.\footnote{Besides \HiBo, no comparable tool with a focus on BSM Higgs searches exists. The code \texttt{Lilith}~\cite{Bernon:2015hsa,Kraml:2019sis} is similar to \HiSi. It, however, at present only includes partial LHC Run-2 results (e.g., no ATLAS results with full Run-2 luminosity are included).}
Both codes, which were written using \texttt{Fortran}, have been developed and extended for roughly one decade.

In this paper, we present a complete rewrite of \HiBo and \HiSi in modern \cpp. \HiBo and \HiSi are now parts of the package \HiTo which also contains \HiPr as a third subpackage facilitating for the user the task of providing theory predictions for the production and decay rates of BSM scalar bosons from the model input. The new setup allows for an easy implementation of new searches and measurements and provides simple-to-use \cpp, \py, and \mat interfaces. It also contains new features like the implementation of non-resonant di-Higgs boson searches, the support for doubly-charged Higgs bosons, or \cp-sensitive coupling measurements. In this paper, we provide a description of the updated codes as well as the newly implemented search limits and rate measurements, and we illustrate the application to several physics examples, demonstrating these new features.

The paper is structured as follows. In \cref{sec:higgstools}, we discuss the \HiTo framework containing the three subpackages \HiPr, \HiBo, and \HiSi. Instructions on how to use this framework are given in \cref{sec:interfaces}. In \cref{sec:examples}, we present several physics examples demonstrating the features of the \HiTo framework. Our conclusions are provided in \cref{sec:conclusions}.


\section{The \HiTo framework}
\label{sec:higgstools}

The \HiTo framework represents a unification and extension of the codes \HiBo and \HiSi. Moreover, it includes the new sub-library \HiPr handling the user input and providing access to many relevant cross sections and branching ratios.

Correspondingly, the \HiTo package contains three subpackages,
\begin{itemize}
  \item \HiPr for defining the physical model and obtaining theory prediction for production and decay rates,
  \item \HiBo for evaluating bounds from direct searches for scalar particles,
  \item \HiSi for evaluating compatibility with the measurements of the Higgs boson detected at ${\sim 125\gev}$.
\end{itemize}
In the following, we will describe the different subpackages in more detail with a special focus on new features with respect to older versions of \HiBo and \HiSi.


\subsection{\HiPr}
\label{sec:hipr}

\HiPr allows defining the physical model. This means that the user has to specify the scalar content of the model and the properties of each BSM scalar. 

These properties include
\begin{itemize}
  \item the mass and total width (including theoretical mass uncertainties),
  \item the electric charge and \cp character,
  \item the rates of all relevant production and decay modes (at LEP and the LHC).
\end{itemize}
All these properties can be set by the user explicitly. Alternatively, the effective coupling input~\cite{Bechtle:2020pkv} can be used to automatically obtain predictions for the most relevant production and decay modes. Moreover, \texttt{SLHA} files or \HiBo data files~\cite{Bechtle:2020pkv} can be used as input via the \py interface.


\subsubsection{Process definitions}

All relevant direct searches for BSM scalars as well as the measurements of the 
properties of the Higgs boson at $125\gev$ are implemented based on the concept of processes.

\begin{figure}
  \centering
  \begin{subfigure}[t]{.45\linewidth}\centering
    \includegraphics{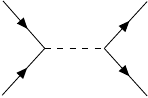}
    \caption{channel}
  \end{subfigure}
  \begin{subfigure}[t]{.45\linewidth}\centering
    \includegraphics{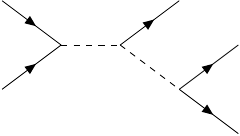}
    \caption{chain decay}
  \end{subfigure}
  \begin{subfigure}[t]{.45\linewidth}\centering
    \includegraphics{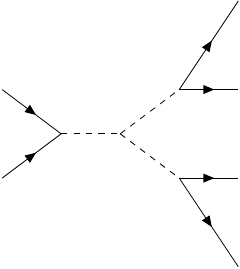}
    \caption{pair decay}
  \end{subfigure}
  \begin{subfigure}[t]{.45\linewidth}\centering
    \includegraphics{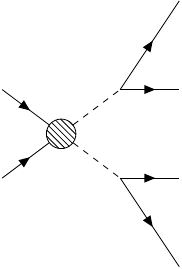}
    \caption{pair production}
  \end{subfigure}
  \caption{Overview of the different process types used within in \HiTo.}
  \label{fig:processes}
\end{figure}

\HiTo currently supports four different types of processes (as depicted in \cref{fig:processes}):
\begin{itemize}
  \item channel processes,
  \item chain decay processes,
  \item pair decay processes,
  \item pair production processes.
\end{itemize}
The channel process is the simplest type of process. It is used for collider processes for which a single (BSM) scalar is produced with specific initial and final states consisting of SM particles. An example process is Higgs production via gluon fusion and the subsequent decay to two photons, $gg \to H \to \gamma\gamma$.

The second type of process is the chain decay process. This process type involves two BSM scalars. The first BSM scalar is produced and then decays to the second BSM scalar and SM particles. The second BSM scalar subsequently decays to SM particles. A typical example is the production of a \cp-odd Higgs boson via gluon fusion followed by a decay to a \cp-even Higgs boson and a $Z$ boson with the second Higgs bosons decaying to bottom quarks, $gg\to A \to Z H \to Z b \bar b$.

The pair decay process extends the chain decay process by a third BSM scalar. The first scalar is produced via a SM initial state and decays into two BSM scalars which subsequently decay to SM particles. One example would be the decay of the SM-like Higgs boson into two light \cp-odd Higgs bosons, which then decay to bottom quarks and two photons, $pp \to h_{125} \to a a \to b\bar b \gamma\gamma$.

It should be noted that \HiTo offers no internal functionality to distinguish between pair decay and pair production processes (or equivalently between resonant and non-resonant pair production). The program relies on the user to provide separate inputs for these two types of processes.

Two BSM scalars can, however, also be produced without originating from the decay of an initial scalar. This possibility is covered by the pair production process which is used for the production of two BSM scalars from a SM initial state and the subsequent decays into SM particles. Typical examples are searches for pair-produced Higgs bosons (performed without the assumption of an initial resonance), e.g.~$pp\to h_1 h_2 \to b\bar b \gamma\gamma$.

The user does not need to provide predictions for every single process. Instead, it is sufficient to input the production cross sections and branching ratios of the relevant scalars. \HiPr will then automatically obtain predictions for every process for which a limit is implemented.

In this procedure, \HiPr will automatically take into account necessary symmetry factors. This is especially relevant for the pair decay and pair production processes, for which two BSM particles $h_i$ and $h_j$ appear in the final state.
Assuming that $h_i$ and $h_j$ decay into the sets of final states $A = \{a_1, a_2, \ldots\}$ and $B = \{b_1, b_2, \ldots\}$, respectively, their combined decay rate is given by
\begin{align}
  \mathrm{Br}(h_i h_j \to A B) =
  \begin{cases}
    \sum_{a\in A}\sum_{b \in B}
      \mathrm{Br}(h_i \to a)\mathrm{Br}(h_j \to b)
    &i\neq j,\\
    \sum_{\{a,b\}\forall a\in A,b\in B} S(\{a,b\})
      \mathrm{Br}(h_i \to a)\mathrm{Br}(h_i \to b)
    &i=j.
  \end{cases}
\end{align}
The sum in the second case runs over all unique unordered pairs $\{a,b\}$, and the symmetry factor $S$ is
\begin{align}
  S(\{a,b\})=
  \begin{cases}
    1 & a=b,\\
    2 & a\neq b.
  \end{cases}
\end{align} 
For example, if $A=\{b b, \tau\tau\}$ and $B=\{bb,\gamma\gamma\}$ the result would be
\begin{align}
  \mathrm{Br}(h_i h_j \to A B) =
  \begin{cases}
    \mathrm{Br}^i_{bb}\mathrm{Br}^j_{bb}
    + \mathrm{Br}^i_{bb}\mathrm{Br}^j_{\gamma\gamma}
    + \mathrm{Br}^i_{\tau\tau}\mathrm{Br}^j_{bb}
    + \mathrm{Br}^i_{\tau\tau}\mathrm{Br}^j_{\gamma\gamma}
    & i\neq j\,\\
    {(\mathrm{Br}^i_{bb})}^2
    + 2\,\mathrm{Br}^i_{bb}\mathrm{Br}^i_{\gamma\gamma}
    + 2\,\mathrm{Br}^i_{\tau\tau}\mathrm{Br}^i_{bb}
    + 2\,\mathrm{Br}^i_{\tau\tau}\mathrm{Br}^i_{\gamma\gamma}
    & i = j,
  \end{cases} 
\end{align}
where $\mathrm{Br}^i_{a}=\mathrm{Br}(h_i\to a)$. Permutations of $h_i$ and $h_j$ when $i\neq j$ are not included at this stage, but are instead accounted for by sums over the corresponding particle clusters $C_1, C_2$,
\begin{align}
  \mathrm{Br}(C_1 C_2 \to A B) =
  \sum_{h_i \in C_1} \sum_{h_j \in C_2}
  \mathrm{Br}(h_i h_j \to A B).
\end{align}
See \cref{sec:clustering} for a detailed discussion of the meaning of particle clusters.

It should be noted that in this branching ratio calculation, electrically charged particle--antiparticle pairs have to be treated as distinct. For this reason, \HiPr currently only allows the implementation of overall neutral final states for pair decay or pair production processes (including final states with two opposite charged BSM scalars).


\subsubsection{Tabulated cross sections and branching ratios}

While production cross section and branching ratio values can completely be provided by the user, \HiPr also provides tabulated reference model predictions for the most common Higgs production and decay channels.

The largest set of predictions is available for scalar bosons with exactly the same couplings as the SM Higgs boson. The tabulated cross section and branching ratio values then only depend on the mass of the new scalar. The numbers for this SM reference model are taken from the website of the LHC Higgs working group (see also \ccite{LHCHiggsCrossSectionWorkingGroup:2016ypw}). For the production cross sections, the following channels are available:
\begin{itemize}
  \item $gg\to H$,
  \item $p p \to b\bar bH$,
  \item $p p \to H + 2 j$ (VBF),
  \item $p p \to WH$,
  \item $p p \to ZH$ (including $gg\to ZH$, $qq\to ZH$, $b\bar b \to ZH$),
  \item $p p \to t\bar t H$,
  \item $p p \to tH$ ($t$ channel + $s$ channel),
  \item $tWH$.
\end{itemize}
All these cross sections are available for scalar masses ranging from $10\gev$ to $3\tev$ and take higher-order QCD corrections into account. For the masses between $120\gev$ and $130\gev$ also predictions including electroweak corrections are available.

Similarly, branching ratio predictions are available for the following decay modes:
\begin{itemize}
  \item $H\to c\bar c, s\bar s, t\bar t, b\bar b$,
  \item $H\to \tau^+\tau^-, \mu^+\mu^-$,
  \item $H \to W^{(*)}W^{(*)}, Z^{(*)}Z^{(*)}, Z\gamma, \gamma\gamma, gg$,
  \item $H\to \text{invisible}$.
\end{itemize}
In \ccite{LHCHiggsCrossSectionWorkingGroup:2016ypw}, numbers are given for masses between $20\gev$ and $900\gev$. We extended this mass range to the interval $1-1000\gev$ using \texttt{HDecay} (version 6.61)~\cite{Djouadi:1997yw,Djouadi:2018xqq}. In addition, we added predictions for the $H\to s\bar s$ decay channel also using \texttt{HDecay}. Predictions including electroweak corrections are available in the mass range $[120,130]\gev$.

In addition to these predictions for scalars with SM-like couplings, \HiPr also implements predictions for scalars with a non-SM-like coupling structure (or non-SM-like quantum numbers). These are available for the  most relevant production modes and encode the dependence on the most relevant couplings. On the other hand, effects from other BSM particles appearing in the production process at the tree or loop level are not taken into account. These cross section predictions can either be directly accessed or automatically used by employing the effective coupling input.

\begin{table}\centering
  \begin{tabular}{cccc}
  \hline
  prod.\ channel             & coupling dep.                 & mass range [GeV]                       & source                                    \\
  \hline         
  $ggH$                        & $c_t, \tilde c_t, c_b, \tilde c_b$  & $10-3000$                              & \texttt{SusHi} \\
  $bbH$                        & $c_b, \tilde c_b$                   & $10-3000$                              & resc.\ of SM result \\
  VBF                          & $c_Z, c_W$                          & LHC8: $1-1050$, LHC13: $1-3050$        & \texttt{HAWK} \\
  $t\bar t H$                  & $c_t,\tilde c_t$                    & $25-1000$                              & \texttt{MadGraph}                         \\
  $tH$ ($t$ channel)           & $c_t,\tilde c_t, c_W$               & $25-1000$                              & \texttt{MadGraph}                         \\
  $tWH$                        & $c_t,\tilde c_t, c_W$               & $25-1000$                              & \texttt{MadGraph}                         \\
  $WH$                         & $c_W, c_t$                          & $1-2950$                               & \texttt{vh@nnlo}                          \\
  $qq\to ZH$                   & $c_Z, c_t$                          & $1-5000$                               & \texttt{vh@nnlo}                          \\
  $gg\to ZH$                   & $c_t,c_b,c_Z,\tilde c_t,\tilde c_b$ & $1-5000$                               & \texttt{vh@nnlo}                          \\
  $b\bar b \to ZH$             & $c_b$                               & $1-5000$                               & \texttt{vh@nnlo}                          \\
  $q_i q_j \to H$              & $c_{q,ij},\tilde c_{q,ij}$          & $1-5000$                               & \texttt{vh@nnlo}                          \\
  $q_i q_j \to H^\pm$          & $c_{qL,ij}, c_{qR,ij}$              & $200-1150$                             & \ccite{Bahl:2021yhk}                     \\
  $q_i q_j \to H + \gamma$     & $c_{q,ij},\tilde c_{q,ij}$          & $200-1150$                             & \ccite{Bahl:2021yhk}                     \\
  $q_i q_j \to H^\pm + \gamma$ & $c_{qL,ij}, c_{qR,ij}$              & $200-1150$                             & \ccite{Bahl:2021yhk}                     \\
  $b\bar b \to ZH$             & $c_b$                               & $200-1150$                             & \ccite{Bahl:2021yhk}                     \\
  $pp\to H^\pm t b$            & $c_{L,tb},c_{R,tb}$                 & $145-2000$                             & \ccite{Degrande:2015vpa,Degrande:2016hyf} \\
  $pp \to H^\pm \phi$          & $c_{H^\pm\phi W^\mp}$               & $m_\phi: 10-500$, $m_{H^\pm}: 100-500$ & \ccite{Bahl:2021str}                      \\
  \hline
  \end{tabular}
  \caption{Overview of cross section predictions available in \HiPr for scalars with a non-SM-like coupling structure.}
  \label{tab:tabulated_XS}
\end{table}

An overview of the implemented cross section predictions for scalars with a non-SM-like coupling structure is given in \cref{tab:tabulated_XS}. The code \texttt{SusHi~1.7.0}~\cite{Harlander:2012pb,Harlander:2016hcx} has been used to derive the cross section for Higgs production via gluon fusion; for the VBF channel, the code \texttt{HAWK~3.0.0}~\cite{Ciccolini:2007jr,Ciccolini:2007ec,Denner:2011id,Denner:2014cla} has been employed; the top-associated Higgs production cross sections have been derived employing the code \texttt{MadGraph5\_aMC@NLO~2.8.2}~\cite{Alwall:2014hca} using the \texttt{MSTW2008LO}~\cite{Martin:2009iq} PDF set assessed via the \texttt{LHAPDF~6.2.3} interface~\cite{Whalley:2005nh}; and for the vector-boson associated Higgs production cross sections, we employed the code \texttt{vh\@nnlo 2.1}~\cite{Brein:2012ne,Harlander:2018yio} (cross-checked with \texttt{MadGraph}~\cite{Bahl:2020wee}).

If not stated otherwise, all cross section predictions in \cref{tab:tabulated_XS} are available only for the 13~TeV LHC. The available cross section predictions will be extended to 13.6~TeV in the near future. In \cref{tab:tabulated_XS}, $c_Z$ and $c_W$ denote the couplings of the scalar to $Z$ and $W$ bosons normalized to the respective coupling of the SM Higgs boson (see \ccite{Bechtle:2020pkv} for more details). Similarly, $c_q$ and $\tilde c_q$ denote the \cp-even and \cp-odd Yukawa couplings to the quark $q$, which are both normalized to the respective \cp-even SM Yukawa coupling (see \ccite{Bechtle:2020pkv} for more details). $c_{q,ij}$ and $\tilde c_{q,ij}$ are used to denote potentially flavor-violating Yukawa couplings to the quarks $q_i$ and $q_j$, for which only the diagonal couplings are normalized to the respective SM values (see \ccite{Bahl:2021yhk} for more details). For charged scalars, $c_{qL,ij}$ and $c_{qR,ij}$ are used to denote the left- and right-handed couplings of the charged scalars to the quarks $q_i$ and $q_j$ (see \ccite{Bahl:2021yhk} for more details). $c_{H^\pm\phi W^\mp}$ denotes the coupling of a charged scalar to a neutral scalar $\phi$ and a $W$ boson (see \ccite{Bahl:2021str}).

All cross sections are calculated automatically by \HiPr if the corresponding effective couplings are used as input. In order to incorporate higher-order corrections in an approximate way, we normalize the fitted cross sections $\sigma_\text{fit}$ by the corresponding cross section evaluated for a Higgs boson with SM-like couplings. The derived number is then multiplied with the respective SM prediction $\sigma_\text{SM}^\text{YR4}$ from \ccite{LHCHiggsCrossSectionWorkingGroup:2016ypw}. This procedure can be summarized in the equation
\begin{align}
  \sigma(m, c_i) = \frac{\sigma_\text{fit}(m, c_i)}{\sigma_\text{fit}(m, c_i^\text{SM})} \sigma_\text{SM}^\text{YR4}(m),
\end{align}
where $c_i$ denotes the set of effective couplings and $c_i^\text{SM}$ the corresponding prediction for a SM Higgs boson. To avoid double-counting, no higher-order SM-like corrections should be included in the calculation of the effective couplings that are used as input for \HiPr.

With respect to the previous implementation of the effective coupling input in \texttt{HiggsBounds-5}, we have improved the handling of heavy Higgs decays to two top quarks. Previously, the decay rates of a \cp-even SM-like Higgs boson given in \ccite{LHCHiggsCrossSectionWorkingGroup:2016ypw} were rescaled by the absolute value of the effective top-Yukawa coupling, $c_t^2 + c_{\tilde t}^2$. This rescaling factor is valid in the limit where the ratio $m_H^2/(4 m_t^2)$ is negligible, where $m_H$ is the mass of the decaying Higgs boson and $m_t$ is the top-quark mass. Going beyond this approximation, the decay rate of a heavy Higgs boson to two top quarks is proportional to
\begin{align}
  \Gamma_{H\to t\bar t} \propto c_t^2 \beta_t^3 + c_{\tilde t}^2 \beta_t \hspace{.3cm}\text{with}\hspace{.3cm}\beta_t = \sqrt{1 - \frac{m_H^2}{4m_t^2}}.
\end{align}
Those different scalings of the parts proportional to the \cp-even and \cp-odd top-Yukawa couplings are now taken into account in \HiPr. In comparison to \texttt{HiggsBounds-5}, this results in increased branching ratios to top quarks for \cp-odd Higgs boson with a mass close to the $t\bar t$ threshold.


\subsection{\HiBo}
\label{sec:hibo}

\HiBo checks the process rates computed by \HiPr based on the input on the considered model provided by the user against a database of experimental limits. For every of these limits, \HiBo performs the following steps:
\begin{itemize}
  \item check which particles in the model are relevant for each role in the process,
  \item find all maximal clusters for each role that fulfil the analysis assumptions,
  \item for all assignments of clusters to the process roles compute the channel rate based on the model predictions provided by \HiPr,
  \item obtain the observed and expected ratios.
\end{itemize}
Then, the most sensitive limit for each particle is selected based on the highest expected ratio. The parameter point is then regarded as allowed if the observed ratio is smaller than one for the most sensitive limit for each particle. This procedure, which has been adopted in order to allow a well-defined statistical interpretation of the applied limit, is described in more detail in \ccite{Bechtle:2008jh,Bechtle:2011sb,Bechtle:2013wla,Bechtle:2020pkv}.

At the moment, the \HiBo limit database contains 258 different experimental limits from LEP and the LHC.


\subsubsection{Limit types}

The main new feature of the new \HiBo implementation, \HiBo{\tt -6}, is a much easier way to incorporate new experimental limits. The whole information about every limit is now encoded in a \texttt{json} file.\footnote{This includes non-trivial acceptance functions as for example used in \ccite{Bahl:2021yhk}.} In the initialization step of \HiBo, a user-specified set of these \texttt{json} files is read-in
and then used for the limit setting.\footnote{For a detailed description of the file format, we refer to \ccite{HiToOnlineDoc_limit_implementation}} For the limit implementation (and evaluation), six different limit types are differentiated:
\begin{itemize}
  \item Channel limit

  A 95\% C.L. limit on the rate of a channel process that only depends on the mass of the particle.

  \textit{Example:} ATLAS search for a heavy Higgs boson produced in association with bottom quarks and decaying into bottom quarks ($pp\to b\bar b h_\text{BSM}\to b\bar b b\bar b$)~\cite{ATLAS:2019tpq}.

  \item Channel width limit

  A 95\% C.L. limit on the rate of a channel process that only depends on the mass and width of the particle.

  \textit{Example:} CMS search for a scalar resonance decaying to a pair of $Z$ bosons ($pp\to h_\text{BSM} \to Z Z$)~\cite{CMS:2018amk}.

  \item Chain decay limit

  A 95\% C.L. limit on the rate of a chain decay process that only depends on the masses of the involved BSM particles.

  \textit{Example:} CMS search for a heavy Higgs boson decaying to a $Z$ boson and a SM-like Higgs boson ($gg\to h_\text{BSM} \to h_{125}Z \to b\bar b \ell^+\ell^-$)~\cite{CMS:2019qcx}.

  \item Pair decay limit

  A 95\% C.L. limit on the rate of a pair decay process that only depends on the masses of the involved BSM particles.

  \textit{Example:} CMS search for a heavy Higgs boson decaying to two SM-like Higgs bosons ($pp\to h_\text{BSM} \to h_{125}h_{125} \to b\bar b \tau^+\tau^-$)~\cite{CMS:2017hea}.

  \item Pair production limit

  A 95\% C.L. limit on the rate of a pair production process that only depends on the masses of the involved BSM particles.

  \textit{Example:} LEP search for pair production of two Higgs bosons ($e^+e^-\to h_1 h_2 \to b \bar b \tau^+\tau^-$)~\cite{ALEPH:2006tnd}.

  \item Likelihood limit

  A limit expressed as a likelihood profile on the rate of multiple channel processes and the mass of the particle.

  \textit{Example:} CMS search for heavy Higgs bosons decaying into two tau leptons which are either produced via gluon fusion or in association with bottom quarks~\cite{CMS:2018rmh}.

\end{itemize}
After reading-in the limit database, \HiBo assigns the BSM scalars and their process rates as provided by \HiPr to specific limits.


\subsubsection{Particle clustering}
\label{sec:clustering}

For each specific limit, all scalars which could participate in the corresponding process are identified. As an example, for the channel process $pp \to \phi \to b\bar b$ the role of $\phi$ within the considered model could be played by $h_{1}$ or $h_{2}$. Then, the set $\{h_{1}, h_{2}\}$ is called a particle cluster.

We define a cluster $C$ of particles with masses $m_1,\ldots,m_i$ to be valid if
\begin{align}
  \text{max}(m_1,\ldots,m_i) - \text{min}(m_1,\ldots,m_i) \le r_{\text{abs}} + r_\text{rel}\cdot\text{mean}(m_1,\ldots,m_i) ,
\end{align}
where $r_\text{abs}$ and $r_\text{rel}$ are the absolute and relative experimental resolutions, respectively. These are either given by the experiment or estimated. In the simple case of a channel limit on $pp\to \phi \to b \bar b$, the limit is then evaluated at the rate-weighted mass and the rate-weighted total width of all particles in the cluster. 

This clustering algorithm has been used already in previous versions of \HiBo/\HiSi, and more details can be found in \ccite{Bechtle:2015pma,Bechtle:2020pkv}. The new implementation in \HiTo extends this functionality by forming clusters for processes involving more than one type of BSM scalars (e.g.\ $pp\to\phi_i\to h_{125}\phi_j,h_{125}\to\tau^+\tau^-,\phi_j\to b\bar b$). In this case, clusters for every BSM particle are formed following the steps outlined above. The limit is then evaluated at the rate-weighted masses and total widths of the clusters formed for $\phi_i$ and $\phi_j$. 

As an example, we consider a BSM model containing the \cp-even scalars $h, H, S$ and the \cp-odd scalars $A$, $A_S$ (as e.g.\ in the N2HDM) with the mass hierarchy $m_H \sim m_A > m_S\sim m_{A_S} > m_h$ (with $m_h \simeq 125\gev$). Then, possible decay modes --- assuming \cp conservation --- are $H\to hh, SS, hS, A_S A_S$ and $A \to h A_S, S A_S$. The clustering algorithm would then potentially assign $H$ and $A$ to one cluster used to compute the mass and width of $\phi_i$ as well as $S$ and $A_S$ to one cluster used to compute the mass and width of $\phi_j$.


\subsubsection{Treatment of mass uncertainties}
\label{sec:mass_unc}

A further improvement of \HiBo with respect to previous versions is the handling of mass uncertainties. In many BSM theories, not all scalar masses are input quantities. Instead, the masses can be calculated in terms of the model parameters (see e.g.\ \ccite{Slavich:2020zjv} for a discussion of Higgs mass predictions in the Minimal Supersymmetric extension of the SM). These theoretical predictions are affected by theoretical and parametric uncertainties induced by unknown higher-order corrections and an imprecise knowledge of the input quantities, respectively.

In previous versions of \HiBo, mass uncertainties have been handled by running the \HiBo algorithm multiple times. For a single scalar with mass uncertainty, the \HiBo algorithm was executed three times: once for the central mass value, once for the central mass value minus the mass uncertainty, and once for the central mass value plus the mass uncertainty. In the end, the result of the run with the weakest constraints was returned. In the case of multiple scalars with mass uncertainties, the \HiBo algorithm was executed for all possible combinations of the central masses plus/minus the associated uncertainties resulting in $3^n$ runs with $n$ being the number of scalars with a mass uncertainty.

The new version of \HiBo improves the handling of mass uncertainties in the following way. As explained, \HiBo only applies the limit with the largest expected ratio. In order to determine the mass value at which each limit is evaluated, the mass is varied within the user-given uncertainty range (checking also intermediate values). Then, the mass with the lowest observed ratio is selected. At this point, also the expected ratio used to compare the sensitivity between the different limits is evaluated.

The usefulness of this prescription becomes clear when discussing experimental searches before the Higgs discovery. Many of these searches have found already hints for the later-discovered Higgs boson in the form of significant excesses around 125~GeV. If \HiBo is now used to derive bounds on a Higgs boson with SM-like couplings and for example a mass of 128~GeV with a mass uncertainty of 4~GeV, the \HiBo algorithm chooses to evaluate the pre-Higgs discovery limits at the mass value where the largest excess was observed, for instance at 125~GeV. As a result, the pre-Higgs discovery limits do not exclude this parameter point in accordance with the fact that the Higgs boson of the considered model can be identified with the Higgs boson that has been detected at 125~GeV (the extent to which the properties of the BSM scalar are compatible with the experimental results on the observed Higgs boson can be tested with \HiSi). If instead the mass value with the largest expected ratio were chosen, this could give rise to an evaluation of the limits at the lower or upper boundary of the mass interval leading to an exclusion of the respective parameter point.

\begin{figure}
  \centering
  \includegraphics[width=0.65\textwidth]{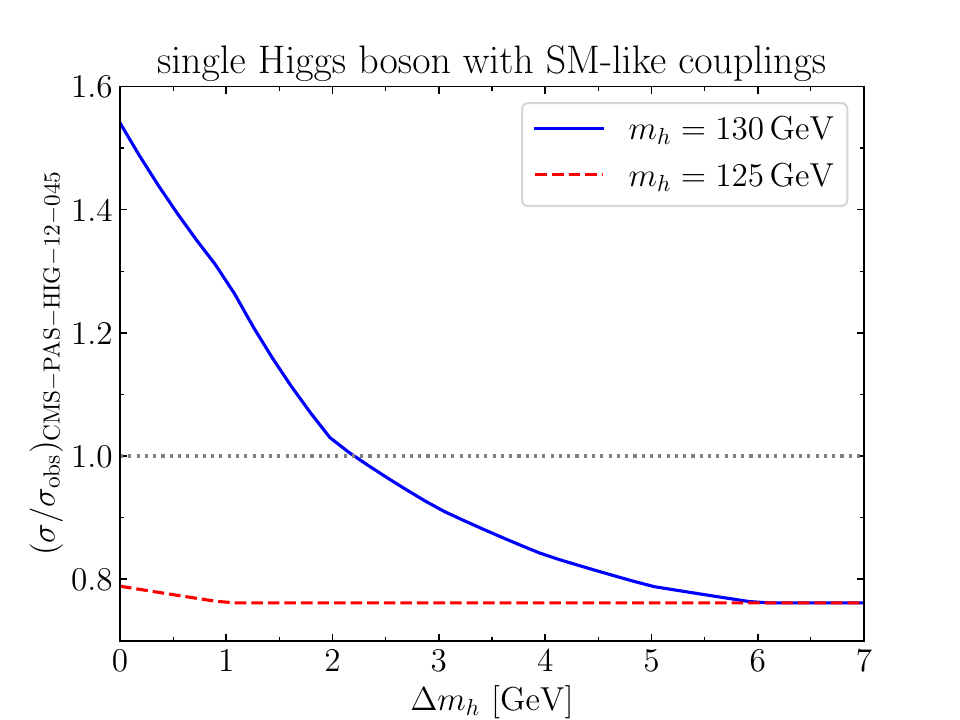}
  \caption{\HiBo observed ratio for the search of \ccite{CMS:2012qwq} derived for a single Higgs boson $h$ with SM-like couplings as a function of its mass uncertainty. The results are shown for two different mass values: $m_h=130\gev$ (blue) and $m_h=125\gev$ (red).}
  \label{fig:obsratio_dm}
\end{figure}

Exemplary results derived using this prescription are shown in \cref{fig:obsratio_dm}. In this Figure, we show the observed ratio for the search of \ccite{CMS:2012qwq} derived for a single Higgs boson $h$ with SM-like couplings as a function of its mass uncertainty. For $m_h=130\gev$, we observe that the observed ratio decreases from $\sim 1.55$ to $\sim 0.75$ as we increase the mass uncertainty from zero to $7\gev$. If instead $m_h=125\gev$, the observed ratio stays essentially constant if increasing the mass uncertainty as expected.


\subsubsection{Higgs pair production limits}
\label{sec:pairprod}

While previous versions of \HiBo had already implemented some searches for resonant Higgs pair production, we significantly extended the scope of implemented searches with the new version, where the included experimental data now additionally comprises the limits of \citeres{CMS:2017rpp,CMS:2018kaz, CMS:2018vjd,CMS:2019noi,CMS:2020jeo, CMS:2021yci,ATLAS:2018fpd,ATLAS:2020azv}. In addition, \HiBo now also implements an extensive list of searches for non-resonant Higgs pair production~\cite{CMS:2017rpp,ATLAS:2018hqk, ATLAS:2018dpp,ATLAS:2018uni,CMS:2018vjd, ATLAS:2018fpd,CMS:2018ipl,ATLAS:2018ili, ATLAS:2019qdc,CMS:2020tkr}.

\HiTo offers no functionality to distinguish between resonant and non-resonant pair production. Therefore, the user has to find a self-defined criterion to decide which part of the considered parameter region should be confronted with resonant or with non-resonant pair-production limits (see e.g.~\ccite{Abouabid:2021yvw}). Then, the input for resonant Higgs pair production can be set by specifying cross section values for pair decay processes and the input for non-resonant Higgs production can be set by providing appropriate values for pair production processes.


\subsubsection{Doubly-charged Higgs bosons}

\begin{table}\centering
  \begin{tabular}{llccc}
  \hline
  Production channel           & decay channel(s)                                                           & $\sqrt{s}$ [TeV] & $\mathcal{L}\;[\fbinv]$ & reference            \\
  \hline         
  $pp\to H^{\pm\pm}H^{\mp\mp}$ & $H^{\pm\pm} \to \ell^\pm \ell^\pm$                                         &                8 &         19.7           & \cite{CMS:2016cpz}   \\
  $pp\to H^{\pm\pm}H^{\mp}$    & $H^{\pm\pm} \to \ell^\pm \ell^{\prime\pm}$, $H^{\mp}\to \ell^\mp \nu_\ell$ &                8 &         19.7           & \cite{CMS:2016cpz}   \\
  $pp\to H^{\pm\pm}H^{\mp\mp}$ & $H^{\pm\pm} \to e^\pm \tau^\pm, \mu^\pm \tau^\pm$                          &                8 &         20.3           & \cite{ATLAS:2014vih} \\
  $pp\to H^{\pm\pm}H^{\mp\mp}$ & $H^{\pm\pm} \to e^\pm e^\pm, e^\pm \mu^\pm, \mu^\pm \mu^\pm$               &                8 &         20.3           & \cite{ATLAS:2014kca} \\
  $pp\to H^{\pm\pm}H^{\mp\mp}$ & $H^{\pm\pm} \to \ell^\pm \ell^\pm$                                         &               13 &         12.9           & \cite{CMS:2017pet}   \\
  $pp\to H^{\pm\pm}H^{\mp}$    & $H^{\pm\pm} \to \ell^\pm \ell^{\prime\pm}$, $H^{\mp}\to \ell^\mp \nu_\ell$ &               13 &         12.9           & \cite{CMS:2017pet}   \\
  $pp\to H^{\pm\pm}H^{\mp\mp}$ & $H^{\pm\pm} \to e^\pm e^\pm, e^\pm \mu^\pm, \mu^\pm \mu^\pm$               &               13 &           36           & \cite{ATLAS:2017xqs} \\
  $pp\to H^{\pm\pm}H^{\mp\mp}$ & $H^{\pm\pm} \to W^\pm W^\pm$                                               &               13 &          139           & \cite{ATLAS:2021jol} \\
  $pp\to H^{\pm\pm}H^{\mp}$    & $H^{\pm\pm} \to W^\pm W^\pm$, $H^{\mp}\to W^{\mp}Z$                        &               13 &          139           & \cite{ATLAS:2021jol} \\
  \hline
  \end{tabular}
  \caption{Overview of implemented doubly charged Higgs searchers (with $\ell,\ell^\prime = e,\mu,\tau$).}
  \label{tab:doubly_charged_Higgs_searches}
\end{table}

As an additional new feature, \HiBo can now also check search limits for doubly-charged Higgs bosons. These appear in triplet or higher multiplet extensions of the SM Higgs sector. An overview of the implemented searches is given in \cref{tab:doubly_charged_Higgs_searches}. While most existing searches concentrate on leptonic final states, with the recent results presented in \ccite{ATLAS:2021jol} also a search with bosonic final states is implemented.


\subsection{\HiSi}

Also \HiSi has been completely reimplemented in modern \cpp. The underlying approach is largely unchanged and has been described in detail in \ccite{Bechtle:2020uwn}. With the reimplementation in modern \cpp, the handling of the different measurement types (i.e., peak-centered observables, mass-centered observables, and STXS measurements) has been unified. As part of this implementation, all individual Higgs mass measurements by ATLAS and CMS have been replaced by a single measurement file based on the PDG combination~\cite{ParticleDataGroup:2022pth}.\footnote{This avoids situations in which a scalar with a theoretical mass uncertainty provides a better fit than a scalar whose mass is exactly at the best-fit point of the PDG combination but not subject to a theoretical mass uncertainty.} At the moment, \HiSi implements 129 individual measurements. Based on these measurements (and the model predictions entered by the user) \HiSi computes a $\chi^2$ value taking into account correlations between the various measurements. This $\chi^2$ value can then be used to test different hypotheses against each other (see recommendations in \ccite{Bechtle:2020uwn}).

In this context, we stress that for models in which the Higgs mass is predicted by the model (e.g.\ in supersymmetric theories) it is crucial to also pass a mass uncertainty to \HiSi in order to avoid a huge $\chi^2$ penalty from the mass measurement. It is futhermore important to provide cross section and branching ratio predictions that are similarly accurate as the state-of-the-art SM predictions (or otherwise use the effective coupling input if adequate).

As a new feature \texttt{HiggsSignals-3} now also contains the implementation of Higgs measurements which are not simple rate measurements but which can also depend on other model parameters. An example is the recent CMS $H\to\tau^+\tau^-$ \cp analysis~\cite{CMS:2021sdq}, which is part of the updated \HiSi dataset. This analysis is targeted at measuring the \cp structure of the tau-Yukawa coupling. The results are presented in dependence on the Higgs production via gluon fusion signal strength, the Higgs production via vector-boson fusion signal strength, and the \cp-violating phase $\phi_\tau$, defined via
\begin{align}
  \tan\phi_\tau = \frac{\tilde c_\tau}{c_\tau} ,
\end{align}
where $c_\tau$ and $\tilde c_\tau$ are the coefficients of the \cp-even and \cp-odd tau-Yukawa coupling (multiplied by the SM tau-Yukawa coupling), respectively. The implementation of such limits depending on the coupling structure of the Higgs boson has not been possible with \texttt{HiggsSignals-2} implying the need to externally evaluate the $\chi^2$. This strategy has e.g.\ been used for the results presented in \ccite{Bahl:2022yrs}. The new \HiSi version allows to fully take such dependencies into account and therefore allows a straightforward implementation of results like the CMS $H\to\tau^+\tau^-$ \cp analysis~\cite{CMS:2021sdq}.


\section{\texttt{C++}, \texttt{Python}, and \texttt{Mathematica} interfaces}
\label{sec:interfaces}

In this Section, we give an overview of the main functionality of the different program parts.  The discussion will not mention (and explain) all available functions and options. Instead, it is aimed at introducing the program flow of \HiTo. For a detailed list of all available functions and options, we refer to the 
online documentation available at
\begin{center}
  \url{https://higgsbounds.gitlab.io/higgstools}.
\end{center}
All shown code snippets are also distributed as parts of complete programs alongside the package.


\subsection{Installation}

\HiTo is available at
\begin{center}
  \url{https://gitlab.com/higgsbounds/higgstools}.
\end{center}
It requires the following software packages:
\begin{itemize}
  \item \texttt{gcc} (at least version \texttt{9}) or \texttt{clang} (at least version \texttt{5}),
  \item \texttt{CMake} (at least version \texttt{3.17}),
  \item for the \py interface: \texttt{python} (at least version \texttt{3.5}) and the corresponding development headers,
  \item for the \mat interface: Wofram \mat.
\end{itemize}
All other dependencies are compile-time only, and are automatically downloaded by \texttt{CMake}. The \HiTo \cpp can be built by running e.g.
\begin{minted}[bgcolor=bg]{bash}
  mkdir build && cd build
  cmake ..
  make
\end{minted}
within the \HiTo directory. 

To build the \py interface, type
\begin{minted}[bgcolor=bg]{bash}
  pip install .
\end{minted}
from within the \HiTo folder (either before or after following the above steps).

To build the \mat executable, use
\begin{minted}[bgcolor=bg]{bash}
  cmake -DHiggsTools_BUILD_MATHEMATICA_INTERFACE=ON ..
\end{minted}
when building the \cpp library (additional information can be found in the online README). The \texttt{MHiggsTools} executable, which can be loaded from within \mat, can then be found in the \texttt{build/wstp} directory.

The collections of limits/measurements for \HiBo and \HiSi are available at \url{https://gitlab.com/higgsbounds/hbdataset} and \url{https://gitlab.com/higgsbounds/hsdataset}.


\subsection{The \cpp and \py interfaces}

First, we explain how to run \HiTo using the \cpp and \py interfaces. The syntax of these interfaces is very similar.

The \cpp libraries are loaded via
\begin{minted}[bgcolor=bg]{cpp}
  #include "Higgs/Bounds.hpp"
  #include "Higgs/Predictions.hpp"
  #include "Higgs/Signals.hpp"

  namespace HP = Higgs::predictions;
\end{minted}
where in the last line a purely optional abbreviation is introduced.

The \HiTo \py package can e.g.\ be loaded via
\begin{minted}[bgcolor=bg]{python}
  import Higgs.predictions as HP
  import Higgs.bounds as HB
  import Higgs.signals as HS
\end{minted}
where again some optional abbreviations are introduced.


\subsubsection{User input via \HiPr}
\label{sec:inputpred}

As a first step, the user has to initialize the \HiPr object e.g.\ by
\begin{minted}[bgcolor=bg]{cpp}
  auto pred = Higgs::Predictions{};
\end{minted}
In \py, one can just write
\begin{minted}[bgcolor=bg]{python}
  pred = Higgs.Predictions()
\end{minted}
This object can then be used to define all relevant scalar bosons via the \texttt{addParticle} function,
\begin{minted}[bgcolor=bg]{cpp}
  auto &h = pred.addParticle(HP::BsmParticle{"h", HP::ECharge::neutral,
                                             HP::CP::even});
\end{minted}
or in \py via
\begin{minted}[bgcolor=bg]{python}
  h = pred.addParticle(HP.BsmParticle("h", "neutral", "even"))
\end{minted}
The properties of the particles can then be defined e.g.\ by
\begin{minted}[bgcolor=bg]{cpp}
  h.setMass(1000);
\end{minted}
or in \py via
\begin{minted}[bgcolor=bg]{python}
  h.setMass(1000)
\end{minted}
Cross section values and partial decay widths can be given via
\begin{minted}[bgcolor=bg]{cpp}
  h.setCxn(HP::Collider::LHC13, HP::Production::ggH, 0.003)
  h.setDecayWidth(HP::Decay::tautau, 0.4)
\end{minted}
or in \py via
\begin{minted}[bgcolor=bg]{python}
  h.setCxn("LHC13", "ggH", 0.003)
  h.setDecayWidth("tautau", 0.4)
\end{minted}
where in the first line we set a cross section of $3\fb$ for $h$ production via gluon fusion at the 13~TeV LHC. In the second line, we set a partial decay width of $400\mev$ for the $h$ decay into two tau leptons. Alternatively, also the total decay width and branching ratios can be set, e.g.\ via
\begin{minted}[bgcolor=bg]{cpp}
  h.setTotalWidth(0.4);
  h.setBr(HP::Decay::tautau, 1);
\end{minted}
or in \py via
\begin{minted}[bgcolor=bg]{python}
  h.setTotalWidth(0.4)
  h.setBr("tautau", 1)
\end{minted}
where in this example we set the total width to $400\mev$ and the branching ratio into tau leptons to 100\%.

As an alternative to providing explicit values for cross sections and decay widths (or branching ratios), the user can also refer to reference models and use the effective coupling input. For example, all couplings of the scalar \texttt{h} can be set to values twice as large as for the SM Higgs boson via
\begin{minted}[bgcolor=bg]{cpp}
  effC = HP::scaledSMlikeEffCouplings(2);
  HP::effectiveCouplingInput(h, effC);
\end{minted}
or in \py via
\begin{minted}[bgcolor=bg]{python}
  effC = HP.scaledSMlikeEffCouplings(2)
  HP.effectiveCouplingInput(h, effC)
\end{minted}
The cross sections and branching ratios of \texttt{h} will then be set automatically using this coupling input. One should note that after setting the effective couplings as given above, one cannot set an additional branching ratio by calling \texttt{h.setBr()}, as this would result in an error due to the fact that the sum of the branching ratios of \texttt{h} would exceed one. Instead, additional decay modes of the scalar can be defined by the user using the function \texttt{h.setDecayWidth()}, in which case internally all previously calculated branching ratios are automatically modified accordingly. We also note that, instead of a global rescaling factor, also all couplings can be set individually (see the example discussed in \refse{sec:charm}). Moreover, for the Higgs--fermion couplings, complex coupling values can be set corresponding to a \cp-even (real part) and a \cp-odd (imaginary part) Yukawa coupling.


\subsubsection{Running \HiBo}

As a first step, one has to initialize \HiBo by
\begin{minted}[bgcolor=bg]{cpp}
  const auto bounds = Higgs::Bounds("/Path/To/HBDataSet");
\end{minted}
or in \py by
\begin{minted}[bgcolor=bg]{python}
  bounds = HB.Bounds("/Path/To/HBDataSet")
\end{minted}
By this command, all limit files in the given folder are read-in and the \HiBo object is created. One can then use this object to check the bounds on a given \HiPr object,
\begin{minted}[bgcolor=bg]{cpp}
  const auto resultHB = bounds(pred);
\end{minted}
or in \py,
\begin{minted}[bgcolor=bg]{python}
  resultHB = bounds(pred)
\end{minted}
The \texttt{result} object will then be either \texttt{True} or \texttt{False} depending on whether the chosen parameter point is allowed or not. More information can be extracted by typing
\begin{minted}[bgcolor=bg]{cpp}
  std::cout << resultHB << std::endl;
\end{minted}
or in \py by typing
\begin{minted}[bgcolor=bg]{python}
  print(resultHB)
\end{minted}
resulting in
\begin{minted}[bgcolor=bg]{text}
  HiggsBounds result: excluded
  particle | obsRatio | expRatio | selected limit description
  ---------|----------|----------|-------------------------------------
      h    |   1.676  |   0.774  | 2d likelihood {LHC13 [ggH>tautau], 
           |          |          | LHC13 [bbH>tautau]} from 2002.12223
           |          |          | (ATLAS 139fb-1, M=(200, 2500))
\end{minted}
as output for the example outlined above (not using the effective coupling input). Alternatively, a list of all selected or applied limits\footnote{All limits are applied to the model predictions. Out of these applied limits, the limit with the highest expected sensitivity is selected for each BSM scalar (see \cref{sec:hibo}).} 
can be obtained via
\begin{minted}[bgcolor=bg]{python}
  resultHB.selectedLimits
\end{minted}
or
\begin{minted}[bgcolor=bg]{python}
  resultHB.appliedLimits
\end{minted}
in either \cpp or \py.


\subsubsection{Running \HiSi}
\label{sec:signals}

In the same way as for \HiBo, the first step in order to use \HiSi is to initialize it by providing the path to the data set folder via
\begin{minted}[bgcolor=bg]{python}
  const auto signals = Higgs::Signals("/Path/To/HSDataSet");
\end{minted}
or in \py via
\begin{minted}[bgcolor=bg]{python}
  signals = HS.Signals("/Path/To/HSDataSet")
\end{minted}
The $\chi^2$ analysis of \HiSi given a \texttt{HiggsPredictions} object as argument can then be invoked via
\begin{minted}[bgcolor=bg]{python}
  auto resultHS = signals(pred);
\end{minted}
or in \py via
\begin{minted}[bgcolor=bg]{python}
  resultHS = signals(pred)
\end{minted}
The return object is the total $\chi^2$ value taking into account the whole data set.

It is also possible to perform the analysis individually for each measurement that is contained in the data set. For instance, in order to obtain the individual $\chi^2$ values for all measurements separately, one can loop over the object \texttt{signals.measurements()} and apply \texttt{signals()} on each element,
\begin{minted}[bgcolor=bg]{python}
  for (const auto &m : signals.measurements()) {
      std::cout << m.reference() << " " << m(pred) << std::endl;
  }
\end{minted}
or in \py via
\begin{minted}[bgcolor=bg]{python}
  for m in signals.measurements():
      print(f"{m.reference()}: {m(pred)}")
\end{minted}
These code snippets return the reference numbers of the experimental measurements and the corresponding individual $\chi^2$ values. Here, it should be noted that the sum of the $\chi^2$ values will be larger than the total $\chi^2$ value, because for obtaining the latter the correlations between the different measurements are taken into account.


\subsection{\mat}

As an alternative to the \cpp/\py interface, \HiTo can also be used via \mat. As a consequence of the different structure of the \texttt{Wolfram} language, the syntax differs from the \cpp/\py interface. 

The Mathematica executable can be loaded via
\begin{minted}[bgcolor=bg]{Mathematica}
  Install["/Path/To/MHiggsTools"];
\end{minted}
This automatically initializes the \HiPr, \HiBo, and \HiSi objects. Particles can then be added via
\begin{minted}[bgcolor=bg]{Mathematica}
  HPAddParticle["H", 1000, "neutral", "even"];
\end{minted}
Their properties can be set e.g.\ via
\begin{minted}[bgcolor=bg]{Mathematica}
  HPSetCxn["H", "LHC13", "ggH", 0.003];

  HPSetDecayWidth["H", "tautau", 0.4];
  (* or *)
  HPSetTotalWidth["H", "LHC13", 0.4];
  HPSetBr["H", "tautau", 1];
\end{minted}
Alternatively the effective coupling input can be used e.g.\ via
\begin{minted}[bgcolor=bg]{Mathematica}
  HPScaledSMlikeEffCouplings["H", 2];
\end{minted}
\HiBo is initialized via
\begin{minted}[bgcolor=bg]{Mathematica}
  HBInitialize["/Path/To/HBDataSet"];
\end{minted}
and run via
\begin{minted}[bgcolor=bg]{Mathematica}
  HBApplyBounds[]
\end{minted}
The applied and selected limits can be assessed via
\begin{minted}[bgcolor=bg]{Mathematica}
  HBGetSelectedBounds[]
  HBGetAppliedBounds[]
\end{minted}
Similarly, \HiSi is initialized via
\begin{minted}[bgcolor=bg]{Mathematica}
  HSInitialize["/Path/To/HSDataSet"];
\end{minted}
and run via
\begin{minted}[bgcolor=bg]{Mathematica}
  HSGetChisq[]
\end{minted}
A list of indidual $\chi^2$ values can be obtained e.g.\ via
\begin{minted}[bgcolor=bg]{Mathematica}
{reference /. #, HSGetChisqMeasurement[id /. #]} & /@ HSListMeasurements[]
\end{minted}
where \texttt{HSListMeasurements[]} returns a list of all loaded measurements.


\section{Examples}
\label{sec:examples}

In this Section, we present some examples using the different components of \HiTo to derive non-trivial constraints on BSM models with an extended or modified Higgs sector.

All these examples use the \py interface. Complete code examples (using also the \cpp and \mat interfaces) are distributed alongside the package. For the \py and \mat scripts also the necessary plotting commands are included.


\subsection{Constraining the charm Yukawa coupling with \HiSi}
\label{sec:charm}

While the couplings of $h_{125}$ to the third-generation fermions have been measured at the LHC at the level of 10\%~\cite{CMS:2018uag,ATLAS:2019nkf}, the couplings to the first- and second-generation fermions are only weakly constrained so far. It is therefore worthwhile to derive indirect constraints on these couplings via the signal-rate measurements of $h_{125}$. We will show here how \HiSi can be utilized to set bounds on the coupling of $h_{125}$ to charm quarks under the assumption that all other couplings of $h_{125}$ are SM-like. This example will also demonstrate the importance of choosing the correct SM reference model for the cross sections of the Higgs boson for the case in which the effective-coupling input is used.

\begin{figure}
    \centering
    \includegraphics[width=0.9\textwidth]{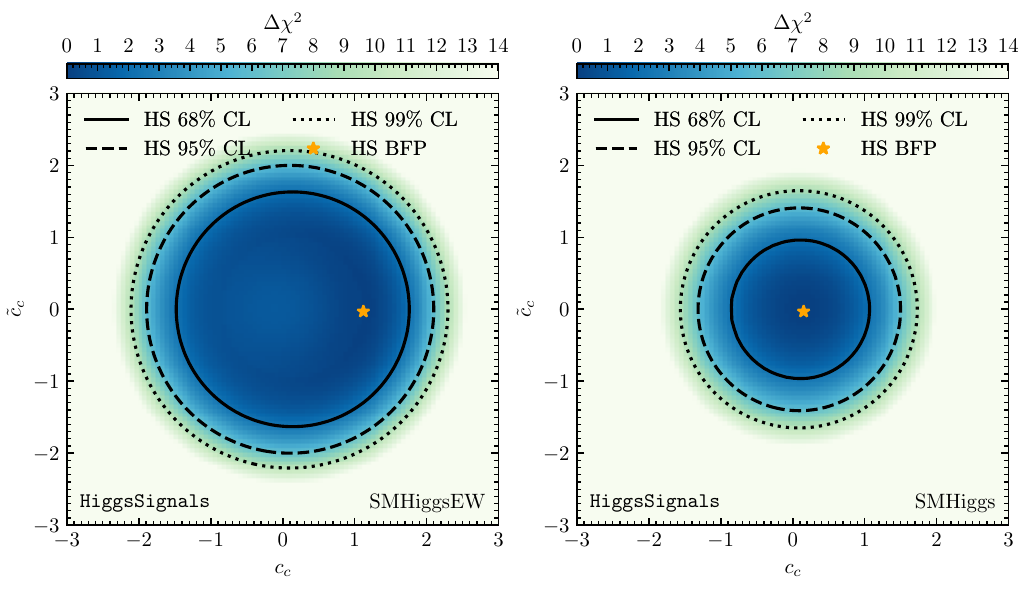}
    \caption{Constraints on the modified coupling of $h_{125}$ to charm quarks in the plane of the coupling modifiers $c_c$ and $\tilde c_c$. In the left plot the reference model \texttt{SMHiggsEW} was used for the \HiSi analysis, whereas the right plot shows the results using the reference model \texttt{SMHiggs}. The orange star in each plot indicates the best fit point of the \HiSi analysis. The SM values are $c_c = 1$ and $\tilde c_c = 0$.}
    \label{fig:cc}
\end{figure}

A state that has the same couplings as a SM Higgs boson except for the charm-quark coupling can be defined in \HiSi in the following way,\footnote{The couplings to first generation fermions are set to their SM values by default.}
\begin{minted}[bgcolor=bg]{python}
  cpls = Higgs.predictions.NeutralEffectiveCouplings()
  cpls.tt = 1
  cpls.bb = 1
  cpls.tautau = 1
  cpls.ss = 1
  cpls.mumu = 1
  cpls.gg = 1
  cpls.ZZ = 1
  cpls.WW = 1
  cpls.gamgam = 1
  cpls.Zgam = 1
  cpls.cc = 0.9 + 1j * 0.1
  Higgs.predictions.effectiveCouplingInput(
      h,
      cpls,
      reference=HP.ReferenceModel.SMHiggsEW)
\end{minted}
As an example, we set here the \cp-even Yukawa coupling to $c_c = 0.9$ and the \cp-odd Yukawa coupling to $\tilde c_c = 0.1$ (times the SM charm-Yukawa coupling), where the latter has to be given as the imaginary component of \texttt{cpls.cc}. Note also that we chose here \texttt{SMHiggsEW} as the reference model in order to utilize the predictions for the cross section of the Higgs boson that include N3LO QCD corrections in the heavy top-quark limit and NLO electroweak corrections.

Following the discussion in \cref{sec:signals}, the $\chi^2$-analysis of \HiSi for the coupling configuration as defined above can now be executed via
\begin{minted}[bgcolor=bg]{python}
  Chisq = signals(pred)
\end{minted}
In the left plot of \cref{fig:cc}, we show the result of the analysis for a scan over both $c_c$ and $\tilde c_c$. The color coding indicates the difference of the $\chi^2$-values with respect to the best-fit point. As expected, the lowest values of $\Delta \chi^2$ are found for $c_c^2 + \tilde c_c^2 = 1$, the region that includes the SM prediction $c_c = 1$ and $\tilde c_c = 0$. The result of \HiSi is different if instead of \texttt{SMHiggsEW} the reference model \texttt{SMHiggs} is chosen in the call of \texttt{effectiveCouplingInput()}. The $\chi^2$ contribution for the option \texttt{SMHiggs} is shown in the right plot of \cref{fig:cc}. One can see that in this case \HiSi finds as the best-fit point the point with vanishing couplings, and the SM would be disfavoured at a confidence level of about $1\sigma$.

The difference between the two plots arises from the fact that the reference model \texttt{SMHiggs} uses the QCD NNLO predictions for the reference cross section, whereas \texttt{SMHiggsEW} uses the N3LO predictions in the heavy top-quark limit~\cite{Aglietti:2004nj,Anastasiou:2014lda,Anastasiou:2015yha,Anastasiou:2016cez}.
Using \texttt{SMHiggs} as reference model
yields a prediction
for the gluon-fusion production cross section of
$\sigma(ggH) = 41.93$~pb,
whereas the \texttt{SMHiggsEW}
reference model yields
$\sigma(ggH) = 48.52$~pb.
If the branching ratios of $h_{125}$ are SM-like,
i.e.~$c_c = 1$ and $\tilde c_c = 0$ in this example,
the signal rates predicted according
to \texttt{SMHiggs} are therefore
slightly smaller than the SM expectation
(which is computed including
the N3LO QCD corrections). If on the other hand
$c_c = 0$, the total decay width of $h_{125}$
is smaller than the SM prediction, such that
the branching ratios in the experimentally observed
decay channels are predicted to be larger
than the SM predictions.
This enhancement of the branching ratios
compensates the smaller prediction for the
gluon-fusion cross section using 
the reference model \texttt{SMHiggs},
and therefore the best-fit $\chi^2$-value is
found for $c_c = \tilde c_c = 0$ in this case.

The comparison between the two plots shows that this kind of analysis is sensitive to QCD corrections beyond the NNLO level in the employed cross section predictions. This example illustrates the importance of choosing the correct reference model. The option \texttt{SMHiggs} is the preferred choice for particles that have a mass comparable to the top-quark mass or larger, whereas for a particle state at $125\gev$ one should use \texttt{SMHiggsEW} in order to include the QCD corrections beyond the NNLO.


\subsection{Sensitivity comparison of resonant \texorpdfstring{$h_{125}$}{h125}-pair production with \HiBo}

\begin{figure}
    \centering
    \includegraphics[width=\textwidth]{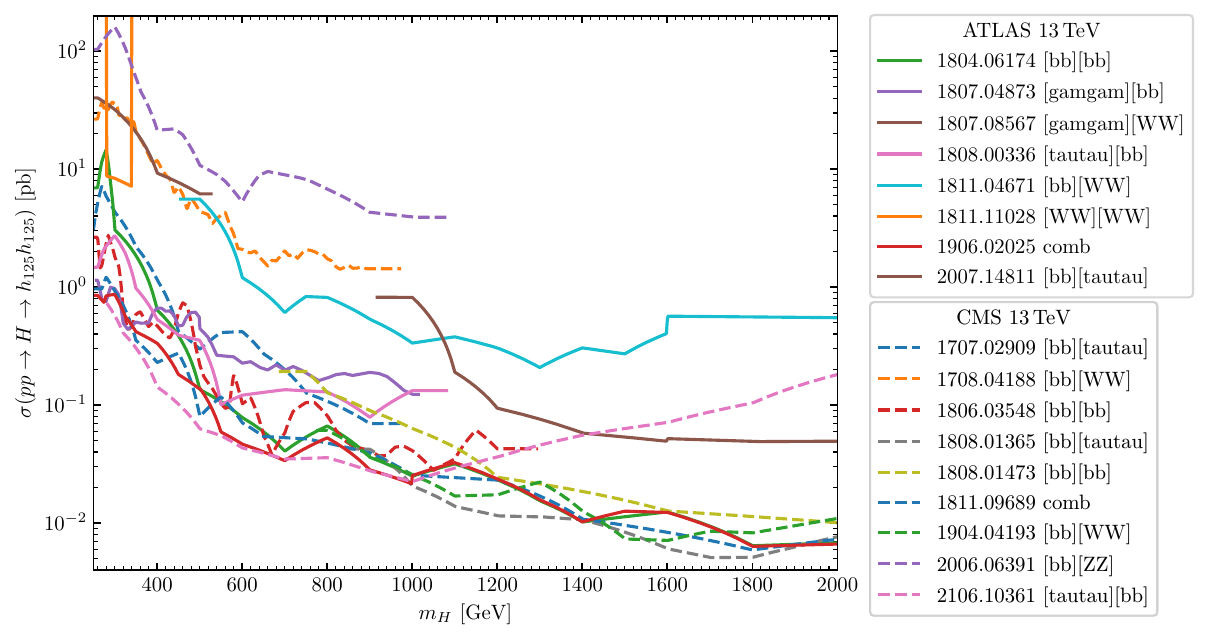}
    \caption{$95\%$ confidence-level cross section limits on the process $pp \to H \to h_{125} h_{125}$ from the experimental searches in various final states. The dashed lines show results from the CMS collaboration, whereas the solid lines show results from the ATLAS collaboration.}
    \label{fig:double}
\end{figure}

As already discussed in \cref{sec:pairprod}, \HiBo{\tt -6}
contains a substantially extended scope of experimental results from searches for resonant pair production of $h_{125}$. In order to illustrate the full extent of the implemented searches, we show in \cref{fig:double} the limits on the cross section $\sigma(pp \to H \to h_{125} h_{125})$ for the different searches as they are currently implemented in the new version of \HiBo. We also show the combined result of both the CMS collaboration (red solid line) and the ATLAS collaboration (blue dashed line) in which the data of various different final states have been included. As discussed in \cref{sec:pairprod}, a large part of the experimental searches shown in \reffi{fig:double} were not yet implemented in the previous \HiBo version. As a result, the new version presented here can give rise to substantially stronger bounds for models in which resonant $h_{125}$-pair production is relevant.

In order to obtain the cross-section limits shown in \cref{fig:double} from their implementation in \HiBo, one can define a SM-like Higgs boson \texttt{h} with a mass of $125\gev$ and a heavy state \texttt{H} with varying mass, and which has a gluon fusion production cross section of $1\,\mathrm{pb}$ and a branching ratio of 1 into $h_{125}$-pairs:
\begin{minted}[bgcolor=bg]{python}
  h = pred.addParticle(HP.NeutralScalar("h", cp="even"))
  H = pred.addParticle(HP.NeutralScalar("H"))
  
  h.setMass(125.09)
  HP.effectiveCouplingInput(h, HP.smLikeEffCouplings)
  H.setDecayWidth("h", "h", 1)
  H.setCxn("LHC13", "ggH", 1)
\end{minted}
Here we defined only one partial width
for the heavy scalar $H$ corresponding
to the decay $H \to h_{125} h_{125}$.
Consequently, independently of the value chosen for this decay width, the corresponding branching ratio is equal to 1. Afterwards, one can call the \HiBo check for different values of the mass of the heavy state and read off the observed ratio for all applied limits that belong to the class of resonant $h_{125}$ pair production:
\begin{minted}[bgcolor=bg]{python}
  masses = np.arange(250, 2001, 10)
  results = {}
  for m in masses:
      H.setMass(m)
      results[m] = [
          l for l in bounds(pred).appliedLimits if
              "H" in l.contributingParticles()]
\end{minted}
The observed ratio is defined as the ratio of predicted cross section and the experimental limit at the $95\%$ confidence level. Since we set the cross section for the process to $1\mathrm{pb}$ independently of the mass of the heavy state, the experimental limit for each mass can thus be obtained by simply calculating the inverse of the observed ratio:
\begin{minted}[bgcolor=bg]{python}
  limits = list({a.limit() for res in results.values() for a in res})
  data = {}
  for lim in limits:
      data[lim.id()] = {
        m: 1 / x.obsRatio() for m, res in results.items() for
            x in res if x.limit() == lim}
\end{minted}
Here the object \texttt{data} contains the
information shown in \cref{fig:double}:
for each applied
experimental search it saves the experimental
limit as a function of the mass of the
scalar \texttt{H}.


\subsection{Constraining the width of the \texorpdfstring{\boldmath{$h_{125}$}}{h125} Higgs boson with \texorpdfstring{\\\HiSi}{HiggsSignals}}
\label{sec:width}

The SM prediction for the total width of the Higgs boson at 125~GeV is $\Gamma^{\rm SM}_{h_{125}} \sim 4\mev$. At a hadron collider, such as the LHC, there is no direct access to the total width of $h_{125}$ under the assumption that $\Gamma_{h_{125}} \ll 1\gev$.\footnote{See \citere{CMS:2022ley} for an indirect measurement of $\Gamma_{h_{125}}$ at the MeV-level via off-shell effects in Higgs boson production. This indirect determination of $\Gamma_{h_{125}}$ relies, however, on several assumptions that are not necessarily fulfilled in BSM scenarios.} As a result, there may be room for new physics that gives rise to modifications of the total width of $h_{125}$ while maintaining values of the measured signal rates of $h_{125}$ close to the SM predictions. The simplest example of such a scenario is a model in which the properties of $h_{125}$ are modified compared to the SM in a twofold way: First, one can assume that there is an additional decay mode of $h_{125}$ that is undetected.\footnote{We note here the distinction between an ``undetected'' decay mode, which cannot be distinguished from the background, and an ``invisible'' decay mode. The latter may very well be detectable because of its characteristic signature of missing energy/momentum in the event.} The branching ratio for this new-physics decay mode is denoted $\mathrm{BR}(h_{125} \to \mathrm{NP})$ in the following, and it gives rise to an enhancement of $\Gamma_{h_{125}}$. In addition, in order to compensate for the suppression of the measured signal rates due to the additional $h \to \mathrm{NP}$ decay mode, one can assume that the couplings of $h_{125}$ to SM particles are enhanced compared to the SM predictions by an overall factor $c_{\rm eff} > 1$.

With \HiSi it is very easy to confront this BSM scenario with the experimental constraints from the LHC. For instance, an enhancement of the couplings by the factor $c_{\rm eff} = 2$ can be set (as already discussed in \cref{sec:inputpred}) with:
\begin{minted}[bgcolor=bg]{python}
  ceff = 2
  cpl = Higgs.predictions.scaledSMlikeEffCouplings(ceff)
  Higgs.predictions.effectiveCouplingInput(
      h,
      cpl,
      reference=Higgs.predictions.ReferenceModel.SMHiggsEW)
\end{minted}
Calling \texttt{effectiveCouplingInput()} automatically invokes the calculation of the partial widths for the decays into SM particles. At the same time, the total width is set to be equal to the sum of all these partial widths. In order to define, for example, $\mathrm{BR}(h_{125} \to \mathrm{NP}) = 0.4$ one can use the function \texttt{setDecayWidth()}.\footnote{One cannot use \texttt{setBR()} at this point, since this would give rise to a runtime error due to the sum of all branching ratios being larger than 1.}
\begin{minted}[bgcolor=bg]{python}
  totalWidthbefore = h.totalWidth()
  branchingRatioNP = 0.4
  partialWidthNP = branchingRatioNP * totalWidthbefore / \
      (1 - branchingRatioNP)
  h.setDecayWidth('NP', partialWidthNP)
\end{minted}
The argument \texttt{'NP'} can be interchanged with any string expression that does not correspond to any of the particle names defined by the user. Now the input is complete and one can call the \HiSi analysis: 
\begin{minted}[bgcolor=bg]{python}
  Chisq = signals(pred)
\end{minted}
Here the objects \texttt{pred} and \texttt{signals} have to be initialized as explained in \cref{sec:inputpred} and \cref{sec:signals}, respectively.

\begin{figure}
    \centering
    \includegraphics[width=0.45\textwidth]{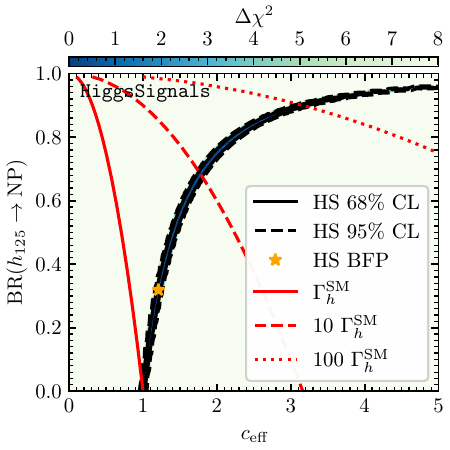}
    \caption{$\chi^2$ result according to the \HiSi analysis for the scenario discussed in \refse{sec:width}. The color coding indicates the value of $\Delta\chi^2$, which is defined relative to the best-fit $\chi^2$-value. The solid and dashed black lines indicate the allowed regions at the $68\%$ and the $95\%$ confidence level, respectively. The orange star indicates the best-fit point. The solid, dashed and dotted red lines indicate where the total width of the Higgs boson has a value of 1, 10 and 100 times the SM prediction, respectively.}
    \label{fig:brnp}
\end{figure}

\cref{fig:brnp} shows the result of the $\chi^2$ analyses when scanning over $0 \leq c_{\rm eff} \leq 5$ and $0 \leq \mathrm{BR}(h_{125} \to \mathrm{NP}) \leq 1$. We find a narrow band in the plane of $c_{\rm eff}$ and $\mathrm{BR}(h_{125} \to \mathrm{NP})$ with low values of $\Delta \chi^2 < 5.99$, where $\Delta \chi^2 = \chi^2 - \chi^2_{\mathrm{best}}$, and $\chi^2_{\mathrm{best}} = \mathrm{min}(\chi^2)$ denotes the best-fit value. We also indicate with red lines the contours for which the total width of $h_{125}$ is equal to 1, 10 or 100 times the SM prediction. The SM point in this plot corresponds to $c_{\rm eff} = 1$ and $\mathrm{BR}(h_{125} \to \mathrm{NP}) = 0$. It has a $\Delta \chi^2 = 0.06$ value relative to the best-fit point that is located at $c_{\rm eff} = 1.21$ and $\mathrm{BR}(h_{125} \to \mathrm{NP}) = 0.32$. As mentioned above,
\HiSi does not compare to the indirect determinations of the total width via off-shell effects, since these analyses are not applicable in a generic fashion to BSM scenarios. For instance, in this example there is a new-physics decay mode for the Higgs boson whose impact on the off-shell effects would have to be taken into account. In fact, currently \HiSi does not contain any direct constraint on the total width of $h_{125}$, such that the region with acceptable values of $\Delta \chi^2$ would extend for $c_{\rm eff} > 5$ until infinity, asymptotically approaching $\mathrm{BR}(h_{125} \to \mathrm{NP}) = 1$. Of course, for total width values in the GeV range the direct search limits from the LHC would apply. Furthermore, the relatively large values of the Higgs couplings that would be needed in this example in order to accommodate  sizable values of $\mathrm{BR}(h_{125} \to \mathrm{NP})$ could also be tested by other experimental constraints.


\subsection{Constraining the 2HDM}
\label{sec:2hdm}

One of the most studied BSM scenarios is the Two-Higgs doublet model (2HDM), which extends the SM by a second SU(2) doublet field~\cite{Lee:1973iz,Kim:1979if} (see \citere{Branco:2011iw} for a review). Assuming \cp conservation, the Higgs sector of the 2HDM consists of two \cp-even states $h$ and $H$, where here we assume that $h$ plays the role of the discovered Higgs boson at $m_h = 125\gev$, a \cp-odd state $A$, and two charged Higgs bosons $H^\pm$, where $m_H$, $m_A$ and $m_{H^\pm}$ are the masses of the BSM Higgs bosons.

In order to avoid flavor-changing neutral currents, a softly broken $Z_2$ symmetry can be introduced, under which one of the Higgs doublets changes the sign, and where $m_{12}^2$ denotes the $Z_2$-breaking mass parameter. There are four different ways of assigning charges of the fermions under the $Z_2$ symmetry, giving rise to the four different Yukawa types of the 2HDM. Each type features a different dependence of the couplings of the Higgs bosons on the angles $\alpha$ and $\tan\beta$, where $\alpha$ is the mixing angle in the \cp-even sector, and $\tan\beta$ is defined as the ratio of the vevs of the \cp-even states.

\begin{figure}[t]
    \centering
    \includegraphics[width=0.40\textwidth]{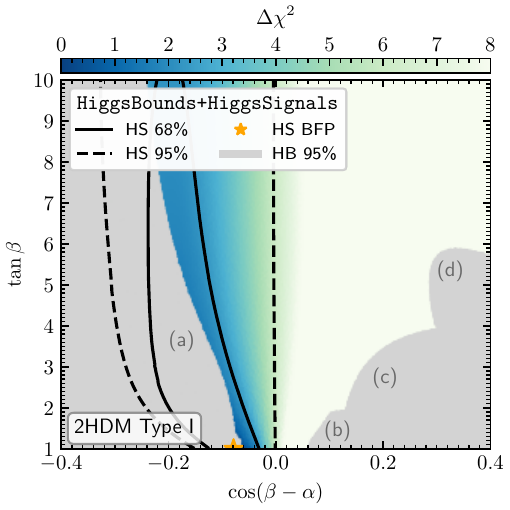}
    \includegraphics[width=0.40\textwidth]{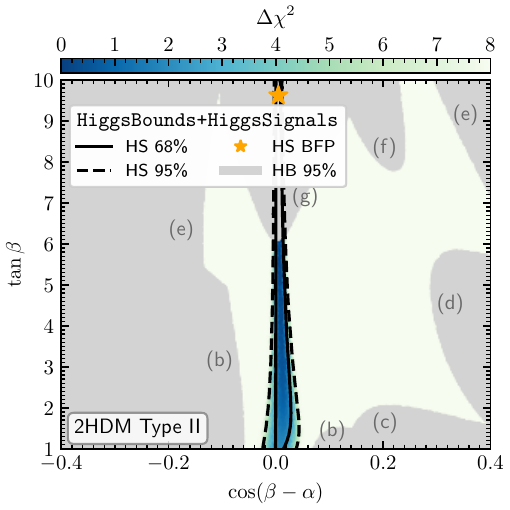}\\
    \includegraphics[width=0.40\textwidth]{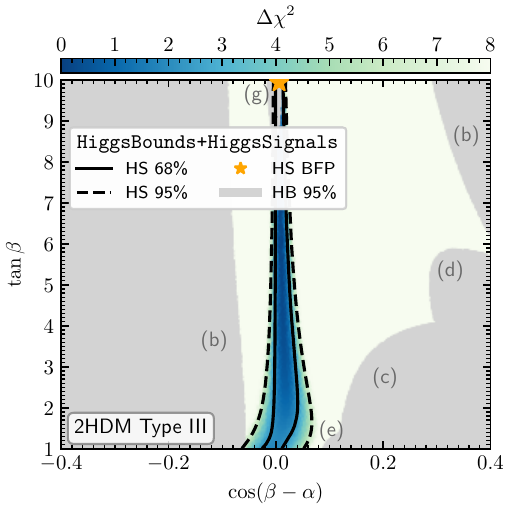}
    \includegraphics[width=0.40\textwidth]{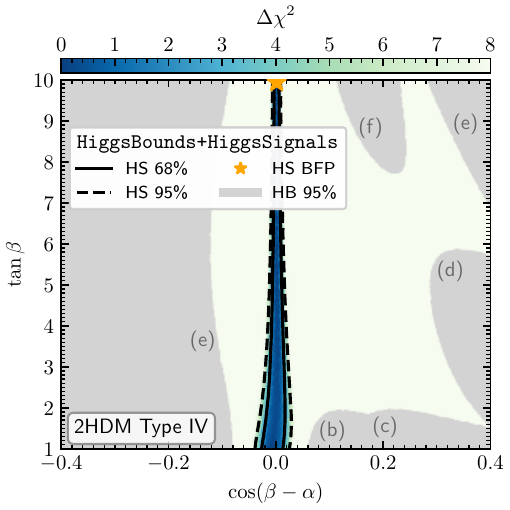}
    \caption{Constraints in the plane of the parameters $\cos ( \beta - \alpha )$ and $\tan\beta$ in the four Yukawa types of the 2HDM for $m_H = m_A = m_{H^\pm} = \sqrt{m_{12}^2 / (\sin\beta \cos\beta)} = 800\gev$. The color coding indicates the value of $\Delta \chi^2$ obtained with \HiSi. The best-fit point with $\Delta \chi^2 = 0$ is indicated with an orange star in each plot. The gray regions are excluded based on the \HiBo result. The gray letters indicate the experimental search responsible for the corresponding exclusion limit. The details of the searches are specified in the text.}
    \label{fig:thdm}
\end{figure}

In \cref{fig:thdm} we show the constraints in the $\cos(\alpha-\beta)$--$\tan\beta$ parameter plane for the four Yukawa types as they result from the $\chi^2$-fit of \HiSi and from the \HiBo analysis.\footnote{A recent detailed analysis of this kind can be found in \ccite{Arco:2022xum}.} The result for $\Delta \chi^2= \chi^2 - \chi^2_{\rm best}$, where $\chi^2_{\rm best}$ denotes the best-fit value, can be obtained by making use of the effective coupling input format, based on the cross sections and branching ratios of $h$ that can be determined with the help of the \HiPr subpackage (see \cref{sec:hipr} for details). The only model information the user has to provide are the effective couplings as functions of $\alpha$ and $\beta$.\footnote{We make the assumption here that since $m_H = m_A = m_{H^\pm} \gg m_h$ the impact of the heavy states $H,A$ and $H^\pm$ on the loop-induced couplings is negligible.} For instance, in order to define the couplings of $h$ to the third-generation fermions in the type~II 2HDM one can write:
\begin{minted}[bgcolor=bg]{python}
  h = pred.addParticle(Higgs.predictions.NeutralScalar("h"))
  cpls_h = Higgs.predictions.NeutralEffectiveCouplings()
  cpls_h.tt = cos(alpha) / sin(beta)
  cpls_h.bb = -sin(alpha) / cos(beta)
  cpls_h.tautau = -sin(alpha) / cos(beta)
  ...
  Higgs.predictions.effectiveCouplingInput(
      h,
      cpls_h,
      reference=Higgs.predictions.ReferenceModel.SMHiggsEW)
\end{minted}
The effective couplings to the vector bosons and the remaining fermions can be set in the same way as before calling \texttt{effectiveCouplingsInput()}. When all effective couplings have been set, the \HiSi $\chi^2$-result is obtained via:
\begin{minted}[bgcolor=bg]{python}
  Chisq = signals(pred)
\end{minted}
It can be observed in \cref{fig:thdm} that for the types~II, III and IV the best-fit point regarding \HiSi (orange star) is
found close to the alignment limit, $\cos(\alpha - \beta) = 0$, in which the properties of $h$ in the 2HDM resemble the ones of a SM Higgs boson. For the Yukawa type~I we find the smallest values of
$\Delta \chi^2$ for slightly negative values
of $\cos(\beta - \alpha)$, which is in agreement with the ATLAS result shown in Fig.~20 of \citere{ATLAS:2021vrm}. With the new version of \HiSi it is very easy to identify the experimental measurement that gives rise to a change of $\chi^2$ in a certain parameter region of a model. In the considered example, for instance for the type~I, one can use \HiSi for two neighbouring parameter points at $\tan\beta=1,\cos(\alpha-\beta)=0$ and at $\tan\beta=1,\cos(\alpha-\beta)=-0.1$ in order to obtain the individual $\chi^2$-values for each implemented measurement by typing (see \cref{sec:signals} for details):
\begin{minted}[bgcolor=bg]{python}
  AllChisq1 = {m.reference(): m(pred1) for m in signals.measurements()}
  AllChisq2 = {m.reference(): m(pred2) for m in signals.measurements()}
  DeltaChisq = {k: AllChisq2[k] - AllChisq1[k] for k in AllChisq1}
\end{minted}
Here \texttt{pred1} and \texttt{pred2} are the \texttt{Higgs.Predictions} objects for the two parameter points that have to be created previously according to the instructions above. In the third line we create a dictionary that contains the differences of the $\chi^2$-values for each experimental measurement. In this example we find that the increase of $\Delta \chi^2$ in the alignment limit of type~I
is mainly driven by the results of \citeres{CMS:2019pyn,ATLAS:2021qou,CMS:2021sdq} which slightly disfavor the alignment limit at the level of about $2\sigma$ each (see also the discussion in \citere{Biekotter:2022ckj}).

In addition to $\Delta \chi^2$, the plots in \cref{fig:thdm} also indicate in gray the parameter regions that are excluded according to the \HiBo analysis. For the considered example, the effective coupling input is not sufficient for calculating the branching ratios for the BSM states.\footnote{The effective coupling input cannot be used for the charged Higgs boson. Moreover, the effects of the charged Higgs boson on the Higgs to di-photon branching ratios are not calculated by \HiPr.} In order to perform this analysis we therefore calculated the cross sections and branching rations with the help of the external software packages \texttt{HDECAY}~\cite{Djouadi:1997yw, Harlander:2013qxa,Djouadi:2018xqq} and \texttt{SuSHi}~\cite{Harlander:2012pb,Harlander:2016hcx} (both called via \texttt{ScannerS}~\cite{Muhlleitner:2020wwk}). The predictions for the cross sections and branching ratios were then provided directly as input to \HiBo. This can be done with the following lines of code, here as an example for the \cp-odd Higgs boson $A$:
\begin{minted}[bgcolor=bg]{python}
  ACP = HP.CP(-1)
  A = pred.addParticle(Higgs.predictions.NeutralScalar("A", ACP))
  A.setMass(...)
  A.setTotalWidth(...)
  A.setBr('tt', ...)
  ...
  A.setBr('Z', 'h', ...)
  A.setCxn('LHC13', 'ggH', ...)
  A.setCxn('LHC13', 'bbH', ...)
\end{minted}
Here the ellipsis in the sixth line indicates additional definitions of branching ratios, whereas the ellipsis in the function arguments represents the numerical values the user has to provide in each case. When the branching ratios and cross sections of $A$, $H$ and $H^\pm$ have been set, the \HiBo analysis is executed by doing:
\begin{minted}[bgcolor=bg]{python}
  res = bounds(pred)
\end{minted}
The desired information of the analysis, for instance the most sensitive channel selected by \HiBo for the state $A$, can be obtained with:
\begin{minted}[bgcolor=bg]{python}
  res.selectedLimits['A']
\end{minted}
The corresponding values for the ratios of predicted cross section divided by expected or observed cross-section limit can be extracted with:
\begin{minted}[bgcolor=bg]{python}
  res.selectedLimits['A'].expRatio()
  res.selectedLimits['A'].obsRatio()
\end{minted}
The information about the other Higgs bosons can be obtained accordingly. The gray regions in the plots in \cref{fig:thdm} are defined by the condition that the \texttt{obsratio} for one of the Higgs bosons is larger than 1. The selected channels responsible for the different excluded regions are the following:
\begin{itemize}
\item[(a)] CMS: $pp \to \phi \to h_{125}
    h_{125} \to bb\gamma\gamma, \
    bb\tau\tau , \ bbbb , \
    bbVV$~\cite{CMS:2018ipl},
\item[(b)] CMS: $pp \to \phi_1 \to h_{125}
    \phi_2 \to bb\tau\tau$~\cite{CMS:2021yci},
\item[(c)] CMS: $pp \to \phi \to Z h_{125}
    \to Z bb$~\cite{CMS:2019qcx},
\item[(d)] ATLAS: $pp \to \phi \to
    WW, \ ZZ, \ WZ$~\cite{ATLAS:2020fry},
\item[(e)] ATLAS: $pp \to \phi \to
    h_{125} h_{125} \to bbbb$~\cite{ATLAS:2018rnh},
\item[(f)] ATLAS: $pp \to \phi \to
    VV, \ V h_{125}$~\cite{ATLAS:2018sbw},
\item[(g)] ATLAS: $pp \to \phi \to
    \tau \tau$~\cite{ATLAS:2020zms},
\end{itemize}
where the letters in \cref{fig:thdm} indicate which limit excludes which parameter region.


\section{Conclusions}
\label{sec:conclusions}

We have presented new versions of the public computer programs \HiBo\ and \HiSi. The program \HiBo tests general BSM models against exclusion limits from LEP and LHC Higgs searches (the limits from Tevatron searches have become less relevant compared to the LHC results and are no longer used). \HiSi confronts the predictions of arbitrary BSM models with the measured mass and rates of the Higgs boson that has been detected at about $125 \gev$. The new versions of \HiBo and \HiSi now use a common interface for the predictions of the various Higgs production and decay rates, provided by the new code \HiPr. The complete suite of codes is provided within the new overarching code \HiTo{\tt -1}.

The description of \HiBo and \HiSi, together with the new code \HiPr, provided in the present paper has focused on the improvements of the functionality and applicability of the programs with respect to the previous versions as given in \citere{Bechtle:2020pkv} and \cite{Bechtle:2020uwn} for \HiBo and \HiSi, respectively. The new code \HiPr, which formerly was partly contained in \HiBo, facilitates the definition of the physical model. The user has to specify the scalar content of the model under consideration: the properties of each scalar of the model, including the mass, total width, charge, \cp character and the rates for all relevant production and decay channels. These properties can be set by the user directly, or alternatively via effective couplings. Concerning the latter option updated tabulated cross sections and branching ratios are included that are then rescaled using the effective coupling input.

Concerning \HiBo, the implemented limits are now classified into six different types, which facilitates the inclusion of new experimental data. These six types comprise {\it (i)} limits for a certain Higgs production and decay chain, {\it (ii)} the same as {\it (i)}, but with a width dependent limit, {\it (iii)} the same as {\it (i)}, but including a longer (Higgs) decay chain, {\it (iv)} limits on Higgs production with a di-Higgs pair decay mode, {\it (v)} limits on di-Higgs production and decay, {\it (vi)} likelihood limits. Also, the algorithm of particle clustering, relevant if several scalars can contribute to a specific limit, has been updated and improved. The set of di-Higgs search channels has been newly implemented (previously only a subset had been considered), as well as limits from searches for doubly charged Higgs bosons. The main update for \HiSi is the extension of the functionality to allow the implementation of Higgs measurements which are not simple rate measurements but which can also depend on other model parameters. This feature has been used in particular for the inclusion of the dedicated \cp analysis by CMS for the decay to $\tau^+\tau^-$, which targets the measurement of the \cp structure of the Higgs coupling to tau leptons. In our description of the codes we have also included detailed information on the \cpp, \py\ and \mat\ interfaces.

We have furthermore discussed in this paper several physics examples with a focus on demonstrating the new functionalities that are provided by \HiTo, together with examples for the commands that have been used to obtain these example applications. The first example concerns constraints on the charm Yukawa coupling that have been obtained with \HiSi. It was demonstrated that the application of the correct Higgs-boson cross-section prediction is crucial to obtain reliable bounds on this Yukawa coupling. In the second physics example the sensitivity of several resonant $h_{125}$-pair production channels that are implemented in \HiBo has been compared. This example demonstrates the effects of the largely extended sample of di-Higgs search limits that are now included into \HiBo. In the third example, updating the previous analysis of \citere{Bechtle:2014ewa} with the latest signal rates that are implemented in \HiSi, a scenario was considered where the total width of the Higgs boson at 125~GeV is enlarged by an undetected decay mode, while the effects of the enlarged width in the signal strength measurements are compensated by a universal scaling of the Higgs-boson couplings. In a final physics example we have demonstrated the combined and complementary power of \HiBo and \HiSi, with the 2HDM as a showcase. In the four Yukawa types of the 2HDM we analyzed the current status of the limits from direct searches (via \HiBo) and from the Higgs-boson rate measurements (via \HiSi) in dependence on the mixing angles, where all heavy Higgs-boson masses were set to $800 \gev$. We showed that for all four Yukawa types significant BSM effects are allowed at the current level of accuracy, where in type~I potentially the largest effects can occur.

The code \HiTo{\tt -1}, containing \HiPr, \HiBo and \HiSi, is available via
\begin{center}
        \url{https://gitlab.com/higgsbounds/higgstools}.
\end{center}


\section*{Acknowledgements}
The authors thank Philip Bechtle, Oliver Brein, Daniel Dercks, Tobias Klingl, Oscar St\r{a}l, Tim Stefaniak, and Karina Williams for collaboration on earlier versions of \HiBo and \HiSi. T.B., C.L., S.P.\ and G.W.~acknowledge support by the Deutsche Forschungsgemeinschaft (DFG, German Research Foundation) under Germany‘s Excellence Strategy – EXC 2121 ``Quantum Universe'' – 390833306. This work has been partially funded by the Deutsche Forschungsgemeinschaft (DFG, German Research Foundation) - 491245950. The work of S.H.\ has received financial support from the grant PID2019-110058GB-C21 funded by MCIN/AEI/10.13039/501100011033 and by ``ERDF A way of making Europe". MEINCOP Spain under contract PID2019-110058GB-C21 and in part by the grant IFT Centro de Excelencia Severo Ochoa CEX2020-001007-S funded by MCIN/AEI/10.13039/501100011033.

\clearpage
\printbibliography

@article{ATLAS:2020zms,
    author = "Aad, Georges and others",
    collaboration = "ATLAS",
    title = "{Search for heavy Higgs bosons decaying into two tau leptons with the ATLAS detector using $pp$ collisions at $\sqrt{s}=13$ TeV}",
    eprint = "2002.12223",
    archivePrefix = "arXiv",
    primaryClass = "hep-ex",
    reportNumber = "CERN-EP-2020-014",
    doi = "10.1103/PhysRevLett.125.051801",
    journal = "Phys. Rev. Lett.",
    volume = "125",
    number = "5",
    pages = "051801",
    year = "2020"
}

@article{ALEPH:2006tnd,
    author = "Schael, S. and others",
    collaboration = "ALEPH, DELPHI, L3, OPAL, LEP Working Group for Higgs Boson Searches",
    title = "{Search for neutral MSSM Higgs bosons at LEP}",
    eprint = "hep-ex/0602042",
    archivePrefix = "arXiv",
    reportNumber = "CERN-PH-EP-2006-001",
    doi = "10.1140/epjc/s2006-02569-7",
    journal = "Eur. Phys. J. C",
    volume = "47",
    pages = "547--587",
    year = "2006"
}

@article{CMS:2017hea,
    author = "Sirunyan, Albert M and others",
    collaboration = "CMS",
    title = "{Search for Higgs boson pair production in events with two bottom quarks and two tau leptons in proton\textendash{}proton collisions at $\sqrt s$ =13TeV}",
    eprint = "1707.02909",
    archivePrefix = "arXiv",
    primaryClass = "hep-ex",
    reportNumber = "CMS-HIG-17-002, CERN-EP-2017-126",
    doi = "10.1016/j.physletb.2018.01.001",
    journal = "Phys. Lett. B",
    volume = "778",
    pages = "101--127",
    year = "2018"
}

@article{CMS:2019qcx,
    author = "Sirunyan, Albert M and others",
    collaboration = "CMS",
    title = "{Search for a heavy pseudoscalar boson decaying to a Z and a Higgs boson at $\sqrt{s} =$ 13 TeV}",
    eprint = "1903.00941",
    archivePrefix = "arXiv",
    primaryClass = "hep-ex",
    reportNumber = "CMS-HIG-18-005, CERN-EP-2018-343",
    doi = "10.1140/epjc/s10052-019-7058-z",
    journal = "Eur. Phys. J. C",
    volume = "79",
    number = "7",
    pages = "564",
    year = "2019"
}

@article{CMS:2018amk,
    author = "Sirunyan, Albert M and others",
    collaboration = "CMS",
    title = "{Search for a new scalar resonance decaying to a pair of Z bosons in proton-proton collisions at $\sqrt{s}=13 $ TeV}",
    eprint = "1804.01939",
    archivePrefix = "arXiv",
    primaryClass = "hep-ex",
    reportNumber = "CMS-HIG-17-012, CERN-EP-2018-009",
    doi = "10.1007/JHEP06(2018)127",
    journal = "JHEP",
    volume = "06",
    pages = "127",
    year = "2018",
    note = "[Erratum: JHEP 03, 128 (2019)]"
}

@article{ATLAS:2019tpq,
    author = "Aad, Georges and others",
    collaboration = "ATLAS",
    title = "{Search for heavy neutral Higgs bosons produced in association with $b$-quarks and decaying into $b$-quarks at $\sqrt{s}=13$ TeV with the ATLAS detector}",
    eprint = "1907.02749",
    archivePrefix = "arXiv",
    primaryClass = "hep-ex",
    reportNumber = "CERN-EP-2019-092",
    doi = "10.1103/PhysRevD.102.032004",
    journal = "Phys. Rev. D",
    volume = "102",
    number = "3",
    pages = "032004",
    year = "2020"
}

@misc{HiToOnlineDoc_limit_implementation,
    title = "HiggsTools online documentation (limit implementation)",
    howpublished = {\url{https://higgsbounds.gitlab.io/higgstools/Datafile.html}}
}

@article{Bahl:2021str,
    author = "Bahl, Henning and Stefaniak, Tim and Wittbrodt, Jonas",
    title = "{The forgotten channels: charged Higgs boson decays to a W$^{\pm}$ and a non-SM-like Higgs boson}",
    eprint = "2103.07484",
    archivePrefix = "arXiv",
    primaryClass = "hep-ph",
    reportNumber = "DESY-21-035, DESY 21-035, LU TP 21-09",
    doi = "10.1007/JHEP06(2021)183",
    journal = "JHEP",
    volume = "06",
    pages = "183",
    year = "2021"
}

@article{Degrande:2015vpa,
    author = "Degrande, Celine and Ubiali, Maria and Wiesemann, Marius and Zaro, Marco",
    title = "{Heavy charged Higgs boson production at the LHC}",
    eprint = "1507.02549",
    archivePrefix = "arXiv",
    primaryClass = "hep-ph",
    reportNumber = "IPPP-15-41, DCPT-15-82, MCNET-15-17, CAVENDISH-HEP-15-04, ZU-TH-18-15",
    doi = "10.1007/JHEP10(2015)145",
    journal = "JHEP",
    volume = "10",
    pages = "145",
    year = "2015"
}

@article{Degrande:2016hyf,
    author = "Degrande, Celine and Frederix, Rikkert and Hirschi, Valentin and Ubiali, Maria and Wiesemann, Marius and Zaro, Marco",
    title = "{Accurate predictions for charged Higgs production: Closing the $m_{H^{\pm}}\sim m_t$ window}",
    eprint = "1607.05291",
    archivePrefix = "arXiv",
    primaryClass = "hep-ph",
    reportNumber = "CAVENDISH-HEP-16-12, IPPP-16-62, SLAC-PUB-16769, TUM-HEP-1055-16, ZH-TH-25-16",
    doi = "10.1016/j.physletb.2017.06.037",
    journal = "Phys. Lett. B",
    volume = "772",
    pages = "87--92",
    year = "2017"
}

@article{LHCHiggsCrossSectionWorkingGroup:2016ypw,
    author = "de Florian, D. and others",
    collaboration = "LHC Higgs Cross Section Working Group",
    title = "{Handbook of LHC Higgs Cross Sections: 4. Deciphering the Nature of the Higgs Sector}",
    eprint = "1610.07922",
    archivePrefix = "arXiv",
    primaryClass = "hep-ph",
    reportNumber = "CERN-2017-002-M, CERN-2017-002",
    doi = "10.23731/CYRM-2017-002",
    volume = "2/2017",
    month = "10",
    year = "2016"
}

@article{Djouadi:1997yw,
    author = "Djouadi, A. and Kalinowski, J. and Spira, M.",
    title = "{HDECAY: A Program for Higgs boson decays in the standard model and its supersymmetric extension}",
    eprint = "hep-ph/9704448",
    archivePrefix = "arXiv",
    reportNumber = "DESY-97-079, IFT-96-29, PM-97-04",
    doi = "10.1016/S0010-4655(97)00123-9",
    journal = "Comput. Phys. Commun.",
    volume = "108",
    pages = "56--74",
    year = "1998"
}

@article{Djouadi:2018xqq,
    author = "Djouadi, Abdelhak and Kalinowski, Jan and Muehlleitner, Margarete and Spira, Michael",
    title = "{HDECAY: Twenty$_{++}$ years after}",
    eprint = "1801.09506",
    archivePrefix = "arXiv",
    primaryClass = "hep-ph",
    reportNumber = "LPT-ORSAY-18-04, CERN-TH-2017-262, LPT-Orsay-18-04, KA-TP-03-2018, PSI-PR-18-02",
    doi = "10.1016/j.cpc.2018.12.010",
    journal = "Comput. Phys. Commun.",
    volume = "238",
    pages = "214--231",
    year = "2019"
}

@article{Bahl:2021yhk,
    author = "Bahl, Henning and Lozano, Victor Martin and Stefaniak, Tim and Wittbrodt, Jonas",
    title = "{Testing Exotic Scalars with HiggsBounds}",
    eprint = "2109.10366",
    archivePrefix = "arXiv",
    primaryClass = "hep-ph",
    reportNumber = "DESY-21-143, LU TP 21-38",
    month = "9",
    year = "2021"
}

@article{ATLAS:2017xqs,
    author = "Aaboud, Morad and others",
    collaboration = "ATLAS",
    title = "{Search for doubly charged Higgs boson production in multi-lepton final states with the ATLAS detector using proton\textendash{}proton collisions at $\sqrt{s}=13\,\text {TeV}$}",
    eprint = "1710.09748",
    archivePrefix = "arXiv",
    primaryClass = "hep-ex",
    reportNumber = "CERN-EP-2017-198",
    doi = "10.1140/epjc/s10052-018-5661-z",
    journal = "Eur. Phys. J. C",
    volume = "78",
    number = "3",
    pages = "199",
    year = "2018"
}

@article{ATLAS:2014kca,
    author = "Aad, Georges and others",
    collaboration = "ATLAS",
    title = "{Search for anomalous production of prompt same-sign lepton pairs and pair-produced doubly charged Higgs bosons with $ \sqrt{s}=8 $ TeV $pp$ collisions using the ATLAS detector}",
    eprint = "1412.0237",
    archivePrefix = "arXiv",
    primaryClass = "hep-ex",
    reportNumber = "CERN-PH-EP-2014-158",
    doi = "10.1007/JHEP03(2015)041",
    journal = "JHEP",
    volume = "03",
    pages = "041",
    year = "2015"
}

@article{ATLAS:2021jol,
    author = "Aad, Georges and others",
    collaboration = "ATLAS",
    title = "{Search for doubly and singly charged Higgs bosons decaying into vector bosons in multi-lepton final states with the ATLAS detector using proton-proton collisions at $\sqrt{s}$ = 13 TeV}",
    eprint = "2101.11961",
    archivePrefix = "arXiv",
    primaryClass = "hep-ex",
    reportNumber = "CERN-EP-2020-240",
    doi = "10.1007/JHEP06(2021)146",
    journal = "JHEP",
    volume = "06",
    pages = "146",
    year = "2021"
}

@article{ATLAS:2014vih,
    author = "Aad, Georges and others",
    collaboration = "ATLAS",
    title = "{Search for new phenomena in events with three or more charged leptons in $pp$ collisions at $\sqrt{s}=8$ TeV with the ATLAS detector}",
    eprint = "1411.2921",
    archivePrefix = "arXiv",
    primaryClass = "hep-ex",
    reportNumber = "CERN-PH-EP-2014-255",
    doi = "10.1007/JHEP08(2015)138",
    journal = "JHEP",
    volume = "08",
    pages = "138",
    year = "2015"
}

@article{CMS:2016cpz,
    collaboration = "CMS",
    title = "{Search for a doubly-charged Higgs boson with $\sqrt{s}=8~\mathrm{TeV}$ $pp$ collisions at the CMS experiment}",
    reportNumber = "CMS-PAS-HIG-14-039",
    year = "2016"
}

@article{CMS:2017pet,
    collaboration = "CMS",
    title = "{A search for doubly-charged Higgs boson production in three and four lepton final states at $\sqrt{s}=13~\mathrm{TeV}$}",
    reportNumber = "CMS-PAS-HIG-16-036",
    year = "2017"
}

@article{Bechtle:2014ewa,
    author = "Bechtle, Philip and Heinemeyer, Sven and St\r{a}l, Oscar and Stefaniak, Tim and Weiglein, Georg",
    title = "{Probing the Standard Model with Higgs signal rates from the Tevatron, the LHC and a future ILC}",
    eprint = "1403.1582",
    archivePrefix = "arXiv",
    primaryClass = "hep-ph",
    reportNumber = "DESY-14-026, BONN-TH-2014-05",
    doi = "10.1007/JHEP11(2014)039",
    journal = "JHEP",
    volume = "11",
    pages = "039",
    year = "2014"
}

@article{CMS:2022ley,
    author = "Tumasyan, Armen and others",
    collaboration = "CMS",
    title = "{Measurement of the Higgs boson width and evidence of its off-shell contributions to ZZ production}",
    eprint = "2202.06923",
    archivePrefix = "arXiv",
    primaryClass = "hep-ex",
    reportNumber = "CMS-HIG-21-013, CERN-EP-2021-272",
    doi = "10.1038/s41567-022-01682-0",
    journal = "Nature Phys.",
    volume = "18",
    number = "11",
    pages = "1329--1334",
    year = "2022"
}

@article{CMS:2018ipl,
    author = "Sirunyan, Albert M and others",
    collaboration = "CMS",
    title = "{Combination of searches for Higgs boson pair production in proton-proton collisions at $\sqrt{s} = $ 13 TeV}",
    eprint = "1811.09689",
    archivePrefix = "arXiv",
    primaryClass = "hep-ex",
    reportNumber = "CMS-HIG-17-030, CERN-EP-2018-292",
    doi = "10.1103/PhysRevLett.122.121803",
    journal = "Phys. Rev. Lett.",
    volume = "122",
    number = "12",
    pages = "121803",
    year = "2019"
}

@article{CMS:2021yci,
    author = "Tumasyan, Armen and others",
    collaboration = "CMS",
    title = "{Search for a heavy Higgs boson decaying into two lighter Higgs bosons in the $\tau\tau$bb final state at 13 TeV}",
    eprint = "2106.10361",
    archivePrefix = "arXiv",
    primaryClass = "hep-ex",
    reportNumber = "CMS-HIG-20-014, CERN-EP-2021-094",
    doi = "10.1007/JHEP11(2021)057",
    journal = "JHEP",
    volume = "11",
    pages = "057",
    year = "2021"
}

@article{ATLAS:2020fry,
    author = "Aad, Georges and others",
    collaboration = "ATLAS",
    title = "{Search for heavy diboson resonances in semileptonic final states in pp collisions at $\sqrt{s}=13$ TeV with the ATLAS detector}",
    eprint = "2004.14636",
    archivePrefix = "arXiv",
    primaryClass = "hep-ex",
    reportNumber = "CERN-EP-2020-049",
    doi = "10.1140/epjc/s10052-020-08554-y",
    journal = "Eur. Phys. J. C",
    volume = "80",
    number = "12",
    pages = "1165",
    year = "2020"
}

@article{ATLAS:2018rnh,
    author = "Aaboud, Morad and others",
    collaboration = "ATLAS",
    title = "{Search for pair production of Higgs bosons in the $b\bar{b}b\bar{b}$ final state using proton-proton collisions at $\sqrt{s} = 13$ TeV with the ATLAS detector}",
    eprint = "1804.06174",
    archivePrefix = "arXiv",
    primaryClass = "hep-ex",
    reportNumber = "CERN-EP-2018-029",
    doi = "10.1007/JHEP01(2019)030",
    journal = "JHEP",
    volume = "01",
    pages = "030",
    year = "2019"
}

@article{ATLAS:2018sbw,
    author = "Aaboud, Morad and others",
    collaboration = "ATLAS",
    title = "{Combination of searches for heavy resonances decaying into bosonic and leptonic final states using 36  fb$^{-1}$ of proton-proton collision data at $\sqrt{s} = 13$ TeV with the ATLAS detector}",
    eprint = "1808.02380",
    archivePrefix = "arXiv",
    primaryClass = "hep-ex",
    reportNumber = "CERN-EP-2018-179",
    doi = "10.1103/PhysRevD.98.052008",
    journal = "Phys. Rev. D",
    volume = "98",
    number = "5",
    pages = "052008",
    year = "2018"
}

@article{Lee:1973iz,
    author = "Lee, T. D.",
    editor = "Feinberg, G.",
    title = "{A Theory of Spontaneous T Violation}",
    doi = "10.1103/PhysRevD.8.1226",
    journal = "Phys. Rev. D",
    volume = "8",
    pages = "1226--1239",
    year = "1973"
}

@article{Kim:1979if,
    author = "Kim, Jihn E.",
    title = "{Weak Interaction Singlet and Strong CP Invariance}",
    reportNumber = "UPR-0120T",
    doi = "10.1103/PhysRevLett.43.103",
    journal = "Phys. Rev. Lett.",
    volume = "43",
    pages = "103",
    year = "1979"
}

@article{Branco:2011iw,
    author = "Branco, G. C. and Ferreira, P. M. and Lavoura, L. and Rebelo, M. N. and Sher, Marc and Silva, Joao P.",
    title = "{Theory and phenomenology of two-Higgs-doublet models}",
    eprint = "1106.0034",
    archivePrefix = "arXiv",
    primaryClass = "hep-ph",
    doi = "10.1016/j.physrep.2012.02.002",
    journal = "Phys. Rept.",
    volume = "516",
    pages = "1--102",
    year = "2012"
}

@article{Harlander:2013qxa,
    author = {Harlander, R. and M\"uhlleitner, M. and Rathsman, J. and Spira, M. and St\r{a}l, O.},
    title = "{Interim recommendations for the evaluation of Higgs production cross sections and branching ratios at the LHC in the Two-Higgs-Doublet Model}",
    eprint = "1312.5571",
    archivePrefix = "arXiv",
    primaryClass = "hep-ph",
    reportNumber = "KA-TP-41-2013, LU-TP-13-44, PSI-PR-13-17, WUB-13-19, LHCHXSWG-2013-001",
    month = "12",
    year = "2013"
}

@article{Harlander:2012pb,
    author = "Harlander, Robert V. and Liebler, Stefan and Mantler, Hendrik",
    title = "{SusHi: A program for the calculation of Higgs production in gluon fusion and bottom-quark annihilation in the Standard Model and the MSSM}",
    eprint = "1212.3249",
    archivePrefix = "arXiv",
    primaryClass = "hep-ph",
    reportNumber = "WUB-12-28, LPN12-134",
    doi = "10.1016/j.cpc.2013.02.006",
    journal = "Comput. Phys. Commun.",
    volume = "184",
    pages = "1605--1617",
    year = "2013"
}

@article{Harlander:2016hcx,
    author = "Harlander, Robert V. and Liebler, Stefan and Mantler, Hendrik",
    title = "{SusHi Bento: Beyond NNLO and the heavy-top limit}",
    eprint = "1605.03190",
    archivePrefix = "arXiv",
    primaryClass = "hep-ph",
    reportNumber = "DESY-16-061, KA-TP-14-2016, TTK-16-14",
    doi = "10.1016/j.cpc.2016.10.015",
    journal = "Comput. Phys. Commun.",
    volume = "212",
    pages = "239--257",
    year = "2017"
}

@article{Muhlleitner:2020wwk,
    author = {M\"uhlleitner, Margarete and Sampaio, Marco O. P. and Santos, Rui and Wittbrodt, Jonas},
    title = "{ScannerS: Parameter Scans in Extended Scalar Sectors}",
    eprint = "2007.02985",
    archivePrefix = "arXiv",
    primaryClass = "hep-ph",
    reportNumber = "KA-TP-05-2020, LU TP 20-38",
    month = "7",
    year = "2020"
}

@article{CMS:2017rpp,
    author = "Sirunyan, Albert M and others",
    collaboration = "CMS",
    title = "{Search for resonant and nonresonant Higgs boson pair production in the $ \mathrm{b}\overline{\mathrm{b}}\mathit{\ell \nu \ell \nu } $ final state in proton-proton collisions at $ \sqrt{s}=13 $ TeV}",
    eprint = "1708.04188",
    archivePrefix = "arXiv",
    primaryClass = "hep-ex",
    reportNumber = "CMS-HIG-17-006, CERN-EP-2017-168",
    doi = "10.1007/JHEP01(2018)054",
    journal = "JHEP",
    volume = "01",
    pages = "054",
    year = "2018"
}

@article{CMS:2018kaz,
    author = "Sirunyan, Albert M and others",
    collaboration = "CMS",
    title = "{Search for heavy resonances decaying into two Higgs bosons or into a Higgs boson and a W or Z boson in proton-proton collisions at 13 TeV}",
    eprint = "1808.01365",
    archivePrefix = "arXiv",
    primaryClass = "hep-ex",
    reportNumber = "CMS-B2G-17-006, CERN-EP-2018-182",
    doi = "10.1007/JHEP01(2019)051",
    journal = "JHEP",
    volume = "01",
    pages = "051",
    year = "2019"
}

@article{CMS:2018vjd,
    author = "Sirunyan, Albert M and others",
    collaboration = "CMS",
    title = "{Search for production of Higgs boson pairs in the four b quark final state using large-area jets in proton-proton collisions at $\sqrt{s}=$ 13 TeV}",
    eprint = "1808.01473",
    archivePrefix = "arXiv",
    primaryClass = "hep-ex",
    reportNumber = "CMS-B2G-17-019, CERN-EP-2018-195",
    doi = "10.1007/JHEP01(2019)040",
    journal = "JHEP",
    volume = "01",
    pages = "040",
    year = "2019"
}

@article{CMS:2019noi,
    author = "Sirunyan, Albert M and others",
    collaboration = "CMS",
    title = "{Search for resonances decaying to a pair of Higgs bosons in the $\mathrm{b\overline{b}q\overline{q}'}\ell\nu$ final state in proton-proton collisions at $\sqrt{s}=$ 13 TeV}",
    eprint = "1904.04193",
    archivePrefix = "arXiv",
    primaryClass = "hep-ex",
    reportNumber = "CMS-B2G-18-008, CERN-EP-2019-056",
    doi = "10.1007/JHEP10(2019)125",
    journal = "JHEP",
    volume = "10",
    pages = "125",
    year = "2019"
}

@article{CMS:2020jeo,
    author = "Sirunyan, Albert M and others",
    collaboration = "CMS",
    title = "{Search for resonant pair production of Higgs bosons in the $bbZZ$ channel in proton-proton collisions at $\sqrt{s}=$ 13 TeV}",
    eprint = "2006.06391",
    archivePrefix = "arXiv",
    primaryClass = "hep-ex",
    reportNumber = "CMS-HIG-18-013, CERN-EP-2020-079",
    doi = "10.1103/PhysRevD.102.032003",
    journal = "Phys. Rev. D",
    volume = "102",
    number = "3",
    pages = "032003",
    year = "2020"
}

@article{ATLAS:2018fpd,
    author = "Aaboud, Morad and others",
    collaboration = "ATLAS",
    title = "{Search for Higgs boson pair production in the $b\bar{b}WW^{*}$ decay mode at $\sqrt{s}=13$ TeV with the ATLAS detector}",
    eprint = "1811.04671",
    archivePrefix = "arXiv",
    primaryClass = "hep-ex",
    reportNumber = "CERN-EP-2018-237",
    doi = "10.1007/JHEP04(2019)092",
    journal = "JHEP",
    volume = "04",
    pages = "092",
    year = "2019"
}

@article{ATLAS:2020azv,
    author = "Aad, Georges and others",
    collaboration = "ATLAS",
    title = "{Reconstruction and identification of boosted di-$\tau$ systems in a search for Higgs boson pairs using 13 TeV proton-proton collision data in ATLAS}",
    eprint = "2007.14811",
    archivePrefix = "arXiv",
    primaryClass = "hep-ex",
    reportNumber = "CERN-EP-2020-118",
    doi = "10.1007/JHEP11(2020)163",
    journal = "JHEP",
    volume = "11",
    pages = "163",
    year = "2020"
}

@article{ATLAS:2019nkf,
    author = "Aad, Georges and others",
    collaboration = "ATLAS",
    title = "{Combined measurements of Higgs boson production and decay using up to $80$ fb$^{-1}$ of proton-proton collision data at $\sqrt{s}=$ 13 TeV collected with the ATLAS experiment}",
    eprint = "1909.02845",
    archivePrefix = "arXiv",
    primaryClass = "hep-ex",
    reportNumber = "CERN-EP-2019-097",
    doi = "10.1103/PhysRevD.101.012002",
    journal = "Phys. Rev. D",
    volume = "101",
    number = "1",
    pages = "012002",
    year = "2020"
}

@article{CMS:2018uag,
    author = "Sirunyan, Albert M and others",
    collaboration = "CMS",
    title = "{Combined measurements of Higgs boson couplings in proton\textendash{}proton collisions at $\sqrt{s}=13\,\text {Te}\text {V} $}",
    eprint = "1809.10733",
    archivePrefix = "arXiv",
    primaryClass = "hep-ex",
    reportNumber = "CMS-HIG-17-031, CERN-EP-2018-263",
    doi = "10.1140/epjc/s10052-019-6909-y",
    journal = "Eur. Phys. J. C",
    volume = "79",
    number = "5",
    pages = "421",
    year = "2019"
}

@article{Ciccolini:2007jr,
    author = "Ciccolini, M. and Denner, Ansgar and Dittmaier, S.",
    title = "{Strong and electroweak corrections to the production of Higgs + 2jets via weak interactions at the LHC}",
    eprint = "0707.0381",
    archivePrefix = "arXiv",
    primaryClass = "hep-ph",
    reportNumber = "MPP-2007-82, PSI-PR-07-03",
    doi = "10.1103/PhysRevLett.99.161803",
    journal = "Phys. Rev. Lett.",
    volume = "99",
    pages = "161803",
    year = "2007"
}

@article{Ciccolini:2007ec,
    author = "Ciccolini, Mariano and Denner, Ansgar and Dittmaier, Stefan",
    title = "{Electroweak and QCD corrections to Higgs production via vector-boson fusion at the LHC}",
    eprint = "0710.4749",
    archivePrefix = "arXiv",
    primaryClass = "hep-ph",
    reportNumber = "MPP-2007-152, PSI-PR-07-06, UWTHPH-2007-26",
    doi = "10.1103/PhysRevD.77.013002",
    journal = "Phys. Rev. D",
    volume = "77",
    pages = "013002",
    year = "2008"
}

@article{Denner:2011id,
    author = "Denner, Ansgar and Dittmaier, Stefan and Kallweit, Stefan and Muck, Alexander",
    title = "{Electroweak corrections to Higgs-strahlung off W/Z bosons at the Tevatron and the LHC with HAWK}",
    eprint = "1112.5142",
    archivePrefix = "arXiv",
    primaryClass = "hep-ph",
    reportNumber = "FR-PHENO-2011-025, PSI-PR-11-04, ZU-TH-29-11, TTK-11-61",
    doi = "10.1007/JHEP03(2012)075",
    journal = "JHEP",
    volume = "03",
    pages = "075",
    year = "2012"
}

@article{Denner:2014cla,
    author = {Denner, Ansgar and Dittmaier, Stefan and Kallweit, Stefan and M\"uck, Alexander},
    title = "{HAWK  2.0: A Monte Carlo program for Higgs production in vector-boson fusion and Higgs strahlung at hadron colliders}",
    eprint = "1412.5390",
    archivePrefix = "arXiv",
    primaryClass = "hep-ph",
    reportNumber = "FR-PHENO-2014-013, MITP-14-101, TTK-14-36",
    doi = "10.1016/j.cpc.2015.04.021",
    journal = "Comput. Phys. Commun.",
    volume = "195",
    pages = "161--171",
    year = "2015"
}

@article{Alwall:2014hca,
      author         = "Alwall, J. and Frederix, R. and Frixione, S. and Hirschi,
                        V. and Maltoni, F. and Mattelaer, O. and Shao, H. -S. and
                        Stelzer, T. and Torrielli, P. and Zaro, M.",
      title          = "{The automated computation of tree-level and
                        next-to-leading order differential cross sections, and
                        their matching to parton shower simulations}",
      journal        = "JHEP",
      volume         = "07",
      year           = "2014",
      pages          = "079",
      doi            = "10.1007/JHEP07(2014)079",
      eprint         = "1405.0301",
      archivePrefix  = "arXiv",
      primaryClass   = "hep-ph",
      reportNumber   = "CERN-PH-TH-2014-064, CP3-14-18, LPN14-066, MCNET-14-09,
                        ZU-TH-14-14",
      SLACcitation   = "%%CITATION = ARXIV:1405.0301;%%"
}

@article{Brein:2012ne,
    author = "Brein, Oliver and Harlander, Robert V. and Zirke, Tom J. E.",
    title = "{vh@nnlo - Higgs Strahlung at hadron colliders}",
    eprint = "1210.5347",
    archivePrefix = "arXiv",
    primaryClass = "hep-ph",
    doi = "10.1016/j.cpc.2012.11.002",
    journal = "Comput. Phys. Commun.",
    volume = "184",
    pages = "998--1003",
    year = "2013"
}

@article{Harlander:2018yio,
    author = "Harlander, Robert V. and Klappert, Jonas and Liebler, Stefan and Simon, Lukas",
    title = "{vh@nnlo-v2: New physics in Higgs Strahlung}",
    eprint = "1802.04817",
    archivePrefix = "arXiv",
    primaryClass = "hep-ph",
    reportNumber = "KA-TP-01-2018, TTK-17-47",
    doi = "10.1007/JHEP05(2018)089",
    journal = "JHEP",
    volume = "05",
    pages = "089",
    year = "2018"
}

@inproceedings{Whalley:2005nh,
    author = "Whalley, M.R. and Bourilkov, D. and Group, R.C.",
    archivePrefix = "arXiv",
    booktitle = "{HERA and the LHC: A Workshop on the implications of HERA for LHC physics. Proceedings, Part B}",
    eprint = "hep-ph/0508110",
    month = "8",
    pages = "575--581",
    title = "{The Les Houches accord PDFs (LHAPDF) and LHAGLUE}",
    year = "2005"
}

@article{Martin:2009iq,
    author = "Martin, A.D. and Stirling, W.J. and Thorne, R.S. and Watt, G.",
    archivePrefix = "arXiv",
    doi = "10.1140/epjc/s10052-009-1072-5",
    eprint = "0901.0002",
    journal = "Eur.\ Phys.\ J.\ C",
    pages = "189--285",
    primaryClass = "hep-ph",
    reportNumber = "IPPP-08-95, DCPT-08-190, CAVENDISH-HEP-08-16",
    title = "{Parton distributions for the LHC}",
    volume = "63",
    year = "2009"
}

@article{Bechtle:2020pkv,
    author = "Bechtle, Philip and Dercks, Daniel and Heinemeyer, Sven and Klingl, Tobias and Stefaniak, Tim and Weiglein, Georg and Wittbrodt, Jonas",
    title = "{HiggsBounds-5: Testing Higgs Sectors in the LHC 13 TeV Era}",
    eprint = "2006.06007",
    archivePrefix = "arXiv",
    primaryClass = "hep-ph",
    reportNumber = "BONN-TH-2020-03, DESY 20-093, DESY-20-093, IFT-UAM/CSIC-20-072, LU 20-27",
    doi = "10.1140/epjc/s10052-020-08557-9",
    journal = "Eur. Phys. J. C",
    volume = "80",
    number = "12",
    pages = "1211",
    year = "2020"
}

@article{Bechtle:2020uwn,
    author = "Bechtle, Philip and Heinemeyer, Sven and Klingl, Tobias and Stefaniak, Tim and Weiglein, Georg and Wittbrodt, Jonas",
    title = "{HiggsSignals-2: Probing new physics with precision Higgs measurements in the LHC 13 TeV era}",
    eprint = "2012.09197",
    archivePrefix = "arXiv",
    primaryClass = "hep-ph",
    reportNumber = "BONN-TH-2020-09, DESY-20-228, DESY 20-228, IFT-UAM/CSIC-20-081, LU TP 20-53",
    doi = "10.1140/epjc/s10052-021-08942-y",
    journal = "Eur. Phys. J. C",
    volume = "81",
    number = "2",
    pages = "145",
    year = "2021"
}

@article{CMS:2021sdq,
    author = "Tumasyan, Armen and others",
    collaboration = "CMS",
    title = "{Analysis of the CP structure of the Yukawa coupling between the Higgs boson and $\tau$ leptons in proton-proton collisions at $\sqrt{s}$ = 13 TeV}",
    eprint = "2110.04836",
    archivePrefix = "arXiv",
    primaryClass = "hep-ex",
    reportNumber = "CMS-HIG-20-006, CERN-EP-2021-189",
    month = "10",
    year = "2021"
}

@article{Bechtle:2015pma,
    author = "Bechtle, Philip and Heinemeyer, Sven and Stal, Oscar and Stefaniak, Tim and Weiglein, Georg",
    title = "{Applying Exclusion Likelihoods from LHC Searches to Extended Higgs Sectors}",
    eprint = "1507.06706",
    archivePrefix = "arXiv",
    primaryClass = "hep-ph",
    reportNumber = "BONN-TH-2015-08, DESY-15-093, SCIPP-15-05",
    doi = "10.1140/epjc/s10052-015-3650-z",
    journal = "Eur. Phys. J. C",
    volume = "75",
    number = "9",
    pages = "421",
    year = "2015"
}

@article{Bechtle:2008jh,
    author = "Bechtle, Philip and Brein, Oliver and Heinemeyer, Sven and Weiglein, Georg and Williams, Karina E.",
    title = "{HiggsBounds: Confronting Arbitrary Higgs Sectors with Exclusion Bounds from LEP and the Tevatron}",
    eprint = "0811.4169",
    archivePrefix = "arXiv",
    primaryClass = "hep-ph",
    reportNumber = "DCPT-08-172, IPPP-08-86, BONN-TH-2008-17",
    doi = "10.1016/j.cpc.2009.09.003",
    journal = "Comput. Phys. Commun.",
    volume = "181",
    pages = "138--167",
    year = "2010"
}

@article{Bechtle:2011sb,
    author = "Bechtle, Philip and Brein, Oliver and Heinemeyer, Sven and Weiglein, Georg and Williams, Karina E.",
    title = "{HiggsBounds 2.0.0: Confronting Neutral and Charged Higgs Sector Predictions with Exclusion Bounds from LEP and the Tevatron}",
    eprint = "1102.1898",
    archivePrefix = "arXiv",
    primaryClass = "hep-ph",
    reportNumber = "FR-PHENO-2011-002, BONN-TH-2011-02, DESY-11-016",
    doi = "10.1016/j.cpc.2011.07.015",
    journal = "Comput. Phys. Commun.",
    volume = "182",
    pages = "2605--2631",
    year = "2011"
}

@article{Bechtle:2013wla,
    author = "Bechtle, Philip and Brein, Oliver and Heinemeyer, Sven and St\r{a}l, Oscar and Stefaniak, Tim and Weiglein, Georg and Williams, Karina E.",
    title = "{$\mathsf{HiggsBounds}-4$: Improved Tests of Extended Higgs Sectors against Exclusion Bounds from LEP, the Tevatron and the LHC}",
    eprint = "1311.0055",
    archivePrefix = "arXiv",
    primaryClass = "hep-ph",
    reportNumber = "BONN-TH-2013-21, DESY-13-110",
    doi = "10.1140/epjc/s10052-013-2693-2",
    journal = "Eur. Phys. J. C",
    volume = "74",
    number = "3",
    pages = "2693",
    year = "2014"
}

@article{Bechtle:2013xfa,
    author = "Bechtle, Philip and Heinemeyer, Sven and St\r{a}l, Oscar and Stefaniak, Tim and Weiglein, Georg",
    title = "{$HiggsSignals$: Confronting arbitrary Higgs sectors with measurements at the Tevatron and the LHC}",
    eprint = "1305.1933",
    archivePrefix = "arXiv",
    primaryClass = "hep-ph",
    reportNumber = "BONN-TH-2013-07, DESY-13-078",
    doi = "10.1140/epjc/s10052-013-2711-4",
    journal = "Eur. Phys. J. C",
    volume = "74",
    number = "2",
    pages = "2711",
    year = "2014"
}

@article{CMS:2012qbp,
    author = "Chatrchyan, Serguei and others",
    collaboration = "CMS",
    title = "{Observation of a New Boson at a Mass of 125 GeV with the CMS Experiment at the LHC}",
    eprint = "1207.7235",
    archivePrefix = "arXiv",
    primaryClass = "hep-ex",
    reportNumber = "CMS-HIG-12-028, CERN-PH-EP-2012-220",
    doi = "10.1016/j.physletb.2012.08.021",
    journal = "Phys. Lett. B",
    volume = "716",
    pages = "30--61",
    year = "2012"
}

@article{ATLAS:2012yve,
    author = "Aad, Georges and others",
    collaboration = "ATLAS",
    title = "{Observation of a new particle in the search for the Standard Model Higgs boson with the ATLAS detector at the LHC}",
    eprint = "1207.7214",
    archivePrefix = "arXiv",
    primaryClass = "hep-ex",
    reportNumber = "CERN-PH-EP-2012-218",
    doi = "10.1016/j.physletb.2012.08.020",
    journal = "Phys. Lett. B",
    volume = "716",
    pages = "1--29",
    year = "2012"
}

@article{ATLAS:2018hqk,
    author = "Aaboud, Morad and others",
    collaboration = "ATLAS",
    title = "{Search for Higgs boson pair production in the $\gamma\gamma WW^{*}$ channel using $pp$ collision data recorded at $\sqrt{s} = 13$ TeV with the ATLAS detector}",
    eprint = "1807.08567",
    archivePrefix = "arXiv",
    primaryClass = "hep-ex",
    reportNumber = "CERN-EP-2018-104",
    doi = "10.1140/epjc/s10052-018-6457-x",
    journal = "Eur. Phys. J. C",
    volume = "78",
    number = "12",
    pages = "1007",
    year = "2018"
}

@article{ATLAS:2018uni,
    author = "Aaboud, Morad and others",
    collaboration = "ATLAS",
    title = "{Search for resonant and non-resonant Higgs boson pair production in the ${b\bar{b}\tau^+\tau^-}$ decay channel in $pp$ collisions at $\sqrt{s}=13$ TeV with the ATLAS detector}",
    eprint = "1808.00336",
    archivePrefix = "arXiv",
    primaryClass = "hep-ex",
    reportNumber = "CERN-EP-2018-164",
    doi = "10.1103/PhysRevLett.121.191801",
    journal = "Phys. Rev. Lett.",
    volume = "121",
    number = "19",
    pages = "191801",
    year = "2018",
    note = "[Erratum: Phys.Rev.Lett. 122, 089901 (2019)]"
}

@article{ATLAS:2018dpp,
    author = "Aaboud, M. and others",
    collaboration = "ATLAS",
    title = "{Search for Higgs boson pair production in the $\gamma\gamma b\bar{b}$ final state with 13 TeV $pp$ collision data collected by the ATLAS experiment}",
    eprint = "1807.04873",
    archivePrefix = "arXiv",
    primaryClass = "hep-ex",
    reportNumber = "CERN-EP-2018-130",
    doi = "10.1007/JHEP11(2018)040",
    journal = "JHEP",
    volume = "11",
    pages = "040",
    year = "2018"
}

@article{ATLAS:2018ili,
    author = "Aaboud, Morad and others",
    collaboration = "ATLAS",
    title = "{Search for Higgs boson pair production in the $WW^{(*)}WW^{(*)}$ decay channel using ATLAS data recorded at $\sqrt{s}=13$ TeV}",
    eprint = "1811.11028",
    archivePrefix = "arXiv",
    primaryClass = "hep-ex",
    reportNumber = "CERN-EP-2018-227",
    doi = "10.1007/JHEP05(2019)124",
    journal = "JHEP",
    volume = "05",
    pages = "124",
    year = "2019"
}

@article{ATLAS:2019qdc,
    author = "Aad, Georges and others",
    collaboration = "ATLAS",
    title = "{Combination of searches for Higgs boson pairs in $pp$ collisions at $\sqrt{s} = $13 TeV with the ATLAS detector}",
    eprint = "1906.02025",
    archivePrefix = "arXiv",
    primaryClass = "hep-ex",
    reportNumber = "CERN-EP-2019-099",
    doi = "10.1016/j.physletb.2019.135103",
    journal = "Phys. Lett. B",
    volume = "800",
    pages = "135103",
    year = "2020"
}

@article{CMS:2020tkr,
    author = "Sirunyan, Albert M and others",
    collaboration = "CMS",
    title = "{Search for nonresonant Higgs boson pair production in final states with two bottom quarks and two photons in proton-proton collisions at $ \sqrt{s} $ = 13 TeV}",
    eprint = "2011.12373",
    archivePrefix = "arXiv",
    primaryClass = "hep-ex",
    reportNumber = "CMS-HIG-19-018, CERN-EP-2020-222",
    doi = "10.1007/JHEP03(2021)257",
    journal = "JHEP",
    volume = "03",
    pages = "257",
    year = "2021"
}

@article{Arco:2022xum,
    author = "Arco, F. and Heinemeyer, S. and Herrero, M. J.",
    title = "{Triple Higgs couplings in the 2HDM: the complete picture}",
    eprint = "2203.12684",
    archivePrefix = "arXiv",
    primaryClass = "hep-ph",
    reportNumber = "IFT--UAM/CSIC-22-032",
    doi = "10.1140/epjc/s10052-022-10485-9",
    journal = "Eur. Phys. J. C",
    volume = "82",
    number = "6",
    pages = "536",
    year = "2022"
}

@article{Bahl:2022yrs,
    author = "Bahl, Henning and Fuchs, Elina and Heinemeyer, Sven and Katzy, Judith and Menen, Marco and Peters, Krisztian and Saimpert, Matthias and Weiglein, Georg",
    title = "{Constraining the ${\mathcal {C}}{\mathcal {P}}$ structure of Higgs-fermion couplings with a global LHC fit, the electron EDM and baryogenesis}",
    eprint = "2202.11753",
    archivePrefix = "arXiv",
    primaryClass = "hep-ph",
    reportNumber = "CERN-TH-2021-231, DESY-22-033, EFI-22-1, IFT--UAM/CSIC--21-148",
    doi = "10.1140/epjc/s10052-022-10528-1",
    journal = "Eur. Phys. J. C",
    volume = "82",
    number = "7",
    pages = "604",
    year = "2022"
}

@article{CMS:2012qwq,
    collaboration = "CMS",
    title = "{Combination of standard model Higgs boson searches and measurements of the properties of the new boson with a mass near 125 GeV}",
    reportNumber = "CMS-PAS-HIG-12-045",
    month = "11",
    year = "2012"
}

@article{Bahl:2020wee,
    author = "Bahl, Henning and Bechtle, Philip and Heinemeyer, Sven and Katzy, Judith and Klingl, Tobias and Peters, Krisztian and Saimpert, Matthias and Stefaniak, Tim and Weiglein, Georg",
    title = "{Indirect $\mathcal{CP}$ probes of the Higgs-top-quark interaction: current LHC constraints and future opportunities}",
    eprint = "2007.08542",
    archivePrefix = "arXiv",
    primaryClass = "hep-ph",
    reportNumber = "DESY-20-102",
    doi = "10.1007/JHEP11(2020)127",
    journal = "JHEP",
    volume = "11",
    pages = "127",
    year = "2020"
}

@article{Abouabid:2021yvw,
    author = {Abouabid, Hamza and Arhrib, Abdesslam and Azevedo, Duarte and Falaki, Jaouad El and Ferreira, Pedro. M. and M\"uhlleitner, Margarete and Santos, Rui},
    title = "{Benchmarking Di-Higgs Production in Various Extended Higgs Sector Models}",
    eprint = "2112.12515",
    archivePrefix = "arXiv",
    primaryClass = "hep-ph",
    month = "12",
    year = "2021"
}

@article{ATLAS:2021vrm,
    collaboration = "ATLAS",
    title = "{Combined measurements of Higgs boson production and decay using up to $139$ fb$^{-1}$ of proton-proton collision data at $\sqrt{s}= 13$ TeV collected with the ATLAS experiment}",
    reportNumber = "ATLAS-CONF-2021-053",
    year = "2021"
}

@article{ATLAS:2021qou,
    author = "Aad, Georges and others",
    collaboration = "ATLAS",
    title = "{Measurement of Higgs boson decay into $b$-quarks in associated production with a top-quark pair in $pp$ collisions at $\sqrt{s}=13$ TeV with the ATLAS detector}",
    eprint = "2111.06712",
    archivePrefix = "arXiv",
    primaryClass = "hep-ex",
    reportNumber = "CERN-EP-2021-202",
    doi = "10.1007/JHEP06(2022)097",
    journal = "JHEP",
    volume = "06",
    pages = "097",
    year = "2022"
}

@article{CMS:2019pyn,
    collaboration = "CMS",
    title = "{Measurement of Higgs boson production and decay to the $\tau\tau$ final state}",
    reportNumber = "CMS-PAS-HIG-18-032",
    year = "2019"
}

@article{Biekotter:2022ckj,
    author = {Biek\"otter, Thomas and Pierre, Mathias},
    title = "{Higgs-boson visible and invisible constraints on hidden sectors}",
    eprint = "2208.05505",
    archivePrefix = "arXiv",
    primaryClass = "hep-ph",
    reportNumber = "DESY-22-128",
    month = "8",
    year = "2022"
}

@article{CMS:2018rmh,
    author = "Sirunyan, Albert M and others",
    collaboration = "CMS",
    title = "{Search for additional neutral MSSM Higgs bosons in the $\tau\tau$ final state in proton-proton collisions at $\sqrt{s}=$ 13 TeV}",
    eprint = "1803.06553",
    archivePrefix = "arXiv",
    primaryClass = "hep-ex",
    reportNumber = "CMS-HIG-17-020, CERN-EP-2018-026",
    doi = "10.1007/JHEP09(2018)007",
    journal = "JHEP",
    volume = "09",
    pages = "007",
    year = "2018"
}

@article{Slavich:2020zjv,
    author = "Slavich, P. and others",
    editor = "Slavich, P. and Heinemeyer, S.",
    title = "{Higgs-mass predictions in the MSSM and beyond}",
    eprint = "2012.15629",
    archivePrefix = "arXiv",
    primaryClass = "hep-ph",
    reportNumber = "DESY 20-229, DESY-20-229, IFT-UAM/CSIC-20-184, FR-PHENO-2020-021, KA-TP-23-2020, MPP-2020-235, P3H-20-086, TTK-20-53, FERMILAB-PUB-21-575-T",
    doi = "10.1140/epjc/s10052-021-09198-2",
    journal = "Eur. Phys. J. C",
    volume = "81",
    number = "5",
    pages = "450",
    year = "2021"
}

@article{Bernon:2015hsa,
    author = "Bernon, Jeremy and Dumont, Beranger",
    title = "{Lilith: a tool for constraining new physics from Higgs measurements}",
    eprint = "1502.04138",
    archivePrefix = "arXiv",
    primaryClass = "hep-ph",
    doi = "10.1140/epjc/s10052-015-3645-9",
    journal = "Eur. Phys. J. C",
    volume = "75",
    number = "9",
    pages = "440",
    year = "2015"
}

@article{Kraml:2019sis,
    author = "Kraml, Sabine and Loc, Tran Quang and Nhung, Dao Thi and Ninh, Le Duc",
    title = "{Constraining new physics from Higgs measurements with Lilith: update to LHC Run 2 results}",
    eprint = "1908.03952",
    archivePrefix = "arXiv",
    primaryClass = "hep-ph",
    reportNumber = "IFIRSE-TH-2019-5",
    doi = "10.21468/SciPostPhys.7.4.052",
    journal = "SciPost Phys.",
    volume = "7",
    number = "4",
    pages = "052",
    year = "2019"
}

@article{ParticleDataGroup:2022pth,
    author = "Workman, R. L. and others",
    collaboration = "Particle Data Group",
    title = "{Review of Particle Physics}",
    doi = "10.1093/ptep/ptac097",
    journal = "PTEP",
    volume = "2022",
    pages = "083C01",
    year = "2022"
}

@article{Aglietti:2004nj,
    author = "Aglietti, U. and Bonciani, R. and Degrassi, G. and Vicini, A.",
    title = "{Two loop light fermion contribution to Higgs production and decays}",
    eprint = "hep-ph/0404071",
    archivePrefix = "arXiv",
    reportNumber = "ROME1-1373-04, FREIBURG-THEP-04-05, RM3-TH-04-06, IFUM-788-FT",
    doi = "10.1016/j.physletb.2004.06.063",
    journal = "Phys. Lett. B",
    volume = "595",
    pages = "432--441",
    year = "2004"
}

@article{Anastasiou:2016cez,
    author = "Anastasiou, Charalampos and Duhr, Claude and Dulat, Falko and Furlan, Elisabetta and Gehrmann, Thomas and Herzog, Franz and Lazopoulos, Achilleas and Mistlberger, Bernhard",
    title = "{High precision determination of the gluon fusion Higgs boson cross-section at the LHC}",
    eprint = "1602.00695",
    archivePrefix = "arXiv",
    primaryClass = "hep-ph",
    reportNumber = "CP3-16-01, ZU-TH-27-15, NIKHEF-2016-004, CERN-TH-2016-006",
    doi = "10.1007/JHEP05(2016)058",
    journal = "JHEP",
    volume = "05",
    pages = "058",
    year = "2016"
}

@article{Anastasiou:2015yha,
    author = "Anastasiou, Charalampos and Duhr, Claude and Dulat, Falko and Furlan, Elisabetta and Herzog, Franz and Mistlberger, Bernhard",
    title = "{Soft expansion of double-real-virtual corrections to Higgs production at N$^{3}$LO}",
    eprint = "1505.04110",
    archivePrefix = "arXiv",
    primaryClass = "hep-ph",
    reportNumber = "CERN-PH-TH-2015-092, CP3-15-11, FERMILAB-PUB-15-089-T, NIKHEF-2015-016",
    doi = "10.1007/JHEP08(2015)051",
    journal = "JHEP",
    volume = "08",
    pages = "051",
    year = "2015"
}

@article{Anastasiou:2014lda,
    author = "Anastasiou, Charalampos and Duhr, Claude and Dulat, Falko and Furlan, Elisabetta and Gehrmann, Thomas and Herzog, Franz and Mistlberger, Bernhard",
    title = "{Higgs Boson Gluon-fusion Production Beyond Threshold in N$^{3}LO$ QCD}",
    eprint = "1411.3584",
    archivePrefix = "arXiv",
    primaryClass = "hep-ph",
    reportNumber = "FERMILAB-PUB-14-461-T",
    doi = "10.1007/JHEP03(2015)091",
    journal = "JHEP",
    volume = "03",
    pages = "091",
    year = "2015"
}

\end{document}